\documentclass{optica-article}
\journal{opticajournal} 
\articletype{Research Article}
\usepackage{lineno}
\usepackage{xspace}
\usepackage{amsmath}
\usepackage{graphicx}
\usepackage{xcolor}
\usepackage{comment}
\usepackage{multirow}
\usepackage{rotating}
\usepackage{lineno}

\newcommand{\etal}{\textit{et al.\@}\xspace}
\newcommand{\exvivo}{\textit{ex vivo}\xspace}

\newcommand{\invivo}{\textit{in vivo}\xspace}

\newcommand{\invitro}{\textit{in vitro}\xspace}

\newcommand{\Enface}{\textit{En face}\xspace}
\newcommand{\enface}{\textit{en face}\xspace}
\newcommand{\um}{\(\muup\)m\xspace}

\newcommand{\ug}{\(\muup\)g\xspace}

\newcommand{\LIV}{\mathrm{LIV}\xspace}

\begin{document}

\title{Label-free intratissue activity imaging of alveolar organoids with dynamic optical coherence tomography}

\author{
	Rion Morishita\authormark{1},
	Toshio Suzuki\authormark{2,3},
	Pradipta Mukherjee\authormark{1},
	Ibrahim Abd El-Sadek\authormark{1,4},
	Yiheng Lim\authormark{1},
	Antonia Lichtenegger\authormark{1,5},
	Shuichi Makita\authormark{1},
	Kiriko Tomita\authormark{1},
	Yuki Yamamoto\authormark{3},
	Tetsuharu Nagamoto\authormark{3},
	and Yoshiaki Yasuno\authormark{1,*}
	
}

\address{
	\authormark{1}Computational Optics Group, University of Tsukuba, Tsukuba, Ibaraki 305-8573, Japan.\\
	\authormark{2}Department of Medical Oncology, Faculty of Medicine, University of Tsukuba, Ibaraki 305-8575, Japan.\\
	\authormark{3}HiLung Inc., Kyoto, Japan.\\
	\authormark{4}Department of Physics, Faculty of Science, Damietta University, New Damietta City 34517, Damietta, Egypt.\\
	\authormark{5} Center for Medical Physics and Biomedical Engineering, Medical University of Vienna, Währinger Gürtel 18-20, 4L, 1090, Vienna, Austria.\\}

\email{\authormark{*}yoshiaki.yasuno@cog-labs.org} 
\homepage{https://optics.bk.tsukuba.ac.jp/COG/}

\begin{abstract}
	An organoid is a three-dimensional (3D) \invitro cell culture emulating human organs.
	We applied 3D dynamic optical coherence tomography (DOCT) to visualize the intratissue and intracellular activities of human induced pluripotent stem cells (hiPSCs)-derived alveolar organoids in normal and fibrosis models.
	3D DOCT data were acquired with an 840-nm spectral domain optical coherence tomography with axial and lateral resolutions of 3.8 \um (in tissue) and 4.9 \um, respectively.
	The DOCT images were obtained by the logarithmic-intensity-variance (LIV) algorithm, which is sensitive to the signal fluctuation magnitude.
	The LIV images revealed cystic structures surrounded by high-LIV borders and mesh-like structures with low LIV.
	The former may be alveoli with a highly dynamics epithelium, while the latter may be fibroblasts.
	The LIV images also demonstrated the abnormal repair of the alveolar epithelium.
\end{abstract}

\section{Introduction}
Recent development of three-dimensional (3D) cell culturing technology enables a variety of organoids \cite{Clevers2016Cell}.
Organoids are 3D \invitro cell culture that closely emulates the structure and function of human organs such as stomach \cite{Bartfeld2015GE}, small intestine \cite{Broda2018NP}, liver \cite{Takebe2013Nature}, and retina \cite{Nakano2012CSC}.
Organoids are widely used for the investigations of genetic activities, cell processes, disease mechanisms, and drug developments \cite{Fei2022Bioengineering}.

An induction method of alveolar epithelial cells from pluripotent stem cells has recently been established \cite{Green2011NB, Longmire2012CSC, Gotoh2014SCR, McCauley2017CSC}.
By exploiting the induction method, alveolar epithelial cells were derived from human induced pluripotent stem cells (hiPSCs), and it enabled the realistic long-term cell culture of alveolar organoids \cite{Yamamoto2017NM}.

Alveolus mainly comprises two types of alveolar epithelial cells, i.e., alveolar type 1 and 2 cells (AT1 and AT2 cells, respectively) \cite{Bertalanffy1955Lancet}.
AT1 cells promote gas exchange, while AT2 cells generate lung surfactant.
It is known that the disorder of alveolar cells causes serious lung diseases \cite{Bueno2014JOCI, Desai2014Nature, Tsuji2006AJOR}.
Therefore, alveolus is an important research target, and alveolar organoids can be used for \invitro investigations.

In addition to hiPSC-derived normal alveolar organoids, hiPSC-derived fibrotic organoids that are made by applying bleomycin, have been reported \cite{Suezawa2021SCR}.
Bleomycin is an anti-cancer drug, however, it is known to induce pulmonary fibrosis, and hence is often used for creating an animal pulmonary fibrosis model \cite{Moeller2008IJM}.
Although such animal models are useful, animal lung cells are anatomically different from human lung cells \cite{Basil2020CSC}, which limits the utility of an animal model.
In contrast, the alveolar organoids are made from human-derived cells, including hiPSC-derived alveolar epithelial cells and human fetal lung fibroblasts.
These organoids can emulate the material metabolism and fibrotic change of the human lung \cite{Yamamoto2017NM, Suezawa2021SCR}. 
Therefore, \invitro alveolar organoids are expected to resolve the limitations of the animal model.

Various imaging techniques have been used to evaluate organoids \cite{Dekkers2019NP}.
The ideal imaging technique should have a field of view (FOV) greater than a millimeter, 3D micrometer-scale resolution, sensitivity to tissue and cell functions, a real-time imaging capability, and low photobleaching and phototoxicity capability (i.e., noninvasiveness) \cite{Fei2022Bioengineering}.

The standard imaging techniques for organoids include bright-field microscopy, fluorescence microscopy, laser scanning confocal microscopy, and electron microscopy.
Bright-field microscopy allows for real-time imaging of living samples, however, it lacks 3D spatial resolution and sensitivity to the tissue/cell functions \cite{Capowski2018Development, Monzel2017SCR, Iefremova2017CR}.
In contrast, fluorescence microscopy can be specific to tissue and cell types, molecules, and gene expressions \cite{Sanderson2014CSHP}.
However, these specificities are achieved using invasive fluorescence agents that cause photobleaching and phototoxicity.
The fluorescence microscopy also lacks 3D spatial resolution \cite{Hoebe2007NB, Frigault2009JOCS}.
Laser scanning confocal microscopy is capable of 3D imaging with approximately 100-\um image penetration, however, it also suffers from photobleaching and phototoxicity \cite{Dekkers2019NP, Sachs2019EMBO}.
Electron microscopy has nanometer-scale resolutions, however it requires a complicated sample preparation, it cannot visualize the inside of the sample, and it does not allow for functional imaging \cite{Sachs2019EMBO, Zewail2010Science}.
Therefore, none of the standard techniques fulfill all of the requirements of organoid imaging.

Dynamic optical coherence tomography (DOCT) is a method to visualize the intratissue and intracellular activities by analyzing the time sequence of OCT images.
First, because the probe is a weak near-infrared light and depth sectioning is achieved by coherence gating, it is non-destructive.
Second, because DOCT uses the intrinsic scattering of the tissue as its contrast source, it is label-free and non-invasive.
Third, because DOCT has more than 1-mm imaging penetration and spatial resolution of a few micrometers, it can visualize thick tissue with cellular scale resolution.
Therefore, DOCT is suitable for organoid evaluation.

Several DOCT methods have been demonstrated to visualize the intratissue and intracellular activities through the fluctuation magnitude \cite{Ling2017LSM, Scholler2019BOE, ElSadek2020BOE}, time-frequency spectrum \cite{Apelian2016BOE,Ling2017LSM,  McLean2017OpEx, Munter2020OL, Leung2020BOE}, and time-correlation property \cite{Lee2012OE, Thouvenin2017IOVS, ElSadek2020BOE} of OCT signals.
DOCT has been applied to a variety of samples including human esophageal and cervical biopsies \cite{Leung2020BOE}, \exvivo mouse or rat organs \cite{Apelian2016BOE, Thouvenin2017IOVS, Scholler2019BOE, Munter2020OL, Mukherjee2021SR, Mukherjee2022BOE}, \invivo zebrafish\cite{Thouvenin2017JOBO}, \invitro cell culture \cite{Park2021BOE}, spheroid \cite{ElSadek2020BOE, ElSadek2021BOE}, mammaria organoid \cite{Oldenburg2015Optica}, and retinal organoid \cite{Scholler2020Light}.
Particularly for pulmonary tissues, Ling \etal demonstrated the DOCT-based visualization and motion analysis of \exvivo ciliated epithelium of human tracheobronchial tissues \cite{Ling2017LSM}.
And McLean \etal demonstrated a principal-component-analysis-based method for the highly specific segmentation of human ciliated epithelium \cite{McLean2017OpEx}.

Among the DOCT techniques, logarithmic-intensity-variance (LIV) contrasts the fluctuation magnitude of the OCT signal intensity \cite{ElSadek2020BOE}, and is expected to be sensitive to the magnitude of intratissue and intracellular motility.
In previous studies, LIV has revealed the functional structures of \exvivo mouse organs \cite{Mukherjee2021SR, Mukherjee2022BOE} and \invitro tumor spheroids \cite{ElSadek2020BOE, ElSadek2021BOE}.

In this paper, we demonstrate LIV-based 3D DOCT imaging of hiPSC-derived alveolar organoids including normal and fibrosis (bleomycin) models.
It is revealed that the alveolar epithelium has high dynamics, exhibits a high LIV.
In addition, some alveolar epithelium shows tessellated patterns of low and high LIV.
We also discuss the histological interpretation of this tessellation, and it is suggested that the tessellation may indicate abnormal repair and remodeling of the epithelium.

\section{Method}
\subsection{Alveolar organoids}
\label{sec:sample}
\begin{figure}
	\centering\includegraphics{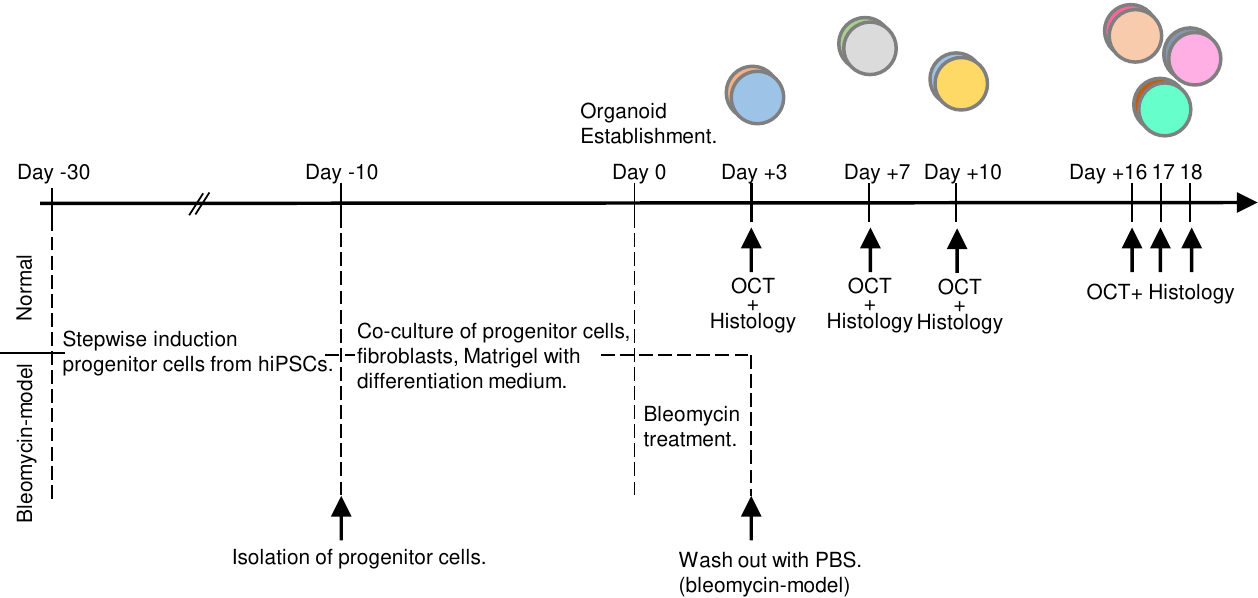}
	\caption{
		The culturing and measurement protocol of hiPSC-induced alveolar organoids including normal and bleomycin models.
		The details of the cultivation and measurement are described in Sections \ref{sec:sample} and \ref{sec:study}, respectively.
		The day of the organoid establishment is set as Day 0.
		The OCT measurements were performed at Days +3, +7, +10, +16, +17, and +18.
		Note that this study is not longitudinal, i.e., the organoids measured at different time points were not the same.
		Each organoid was fixed in formalin after the OCT measurement for histological preparation.}
	\label{fig:timeChart}
\end{figure}

Two types of hiPSC-derived alveolar organoids, including the normal and fibrosis models, were used in this study.
Throughout this manuscript, the fibrosis model is denoted as the bleomycin model.

The alveolar organoids are formed by co-culture of the hiPSC-derived lung progenitor cells and human fetal lung fibroblasts according to the published protocol by Gotoh \etal\cite{Gotoh2014SCR} and Yamamoto \etal\cite{Yamamoto2017NM} with modifications.
The time course of the protocol is shown in Fig.\@ \ref{fig:timeChart} (from Day -30 to Day 0).
NKX2.1+lung progenitor cells were stepwise-induced from hiPSCs (HILC line, HiLung, Kyoto, Japan) using previously reported protocols\cite{Yamamoto2017NM}.
At Day -10 (20 days after starting the stepwise induction), NKX2.1+lung progenitor cells were isolated by cell sorting with a specific surface marker, carboxypeptidase-M (CPM).
After the isolation, from Day -10 to Day 0, the NKX2.1+lung progenitor cells were co-cultured with human lung fetal lung fibroblasts (TIG-1-20, JCRB0501, JCRB Cell Bank) embedded in 50\%Matrigel Growth Factor Reduced (GFR) Basement Membrane Matrix (\#354230, Corning) on a cell culture insert (\#353180, Falcon) with alveolar differentiation medium.
The alveolar differentiation medium was supplemented with dexamethasone (Cat\# D4902, Sigma-Aldrich, MO), KGF (Cat\# 100-19, PeproTech, NJ), 8-Br-cAMP (Cat\# B007, Biolog, CA), and 3-isobutyl 1-methylxanthine (IBMX) (Cat\# 095-03413, FUJIFILM Wako, Osaka, Japan).

For the bleomycin model, after 10 days of cultivation, the alveolar organoid was treated with 10 \ug/mL bleomycin (Nippon Kayaku, Tokyo, Japan) in the medium for 72 hours followed by washing out using phosphate buffered saline (PBS) (Nacalai Tesque, Kyoto, Japan).

\subsection{Study protocol}
\label{sec:study}
Normal and bleomycin-model organoids cultured for 3, 7, 10, 16, 17, and 18 days after the organoid establishment were prepared as shown in Fig.\@ \ref{fig:timeChart} (from Day 0 to Day +18).
Here the Days +16 to +18 are collectively regarded as ``a late time point.’’
So, the measurement interval is shorter in the earlier time and longer in the later time.
This non-uniform interval was selected because the organoids more rapidly change in the earlier time, while they become more stable in later time.
In addition, it was expected that the variation among the individual samples can be large in the late time point.
And hence, the protocol was designed to measure more samples in the late time point (i.e., six samples in total at Days +16 to +18). 

For DOCT measurement, the organoid was transferred from the cell culture insert to a Petri dish with culture medium.
A black mending tape was glued to the bottom of the Petri dish to prevent strong back-reflection from the bottom surface of the dish.
The lateral scan ranges were 6 mm $\times$ 6 mm, 3 mm $\times$ 3 mm, and 1 mm $\times$ 1 mm.
Here the widest FOV is to cover the whole organoid region, and it has been used to select particular regions of interest for the smaller FOV measurements.
The middle FOV is for the primary image analysis described in Section \ref{sec:image analysis}.
And the smallest FOV is for detailed observation of the dynamics and structure of each alveolus, such as shown in Figs.\@ \ref{fig:normal organoid} and \ref{fig:alveoli}.
The details of the scan protocol are described in Section \ref{sec:OCT method}.
After the DOCT measurement, the organoid was fixed in formalin for histology imaging.
The details of the histology imaging are described in Section \ref{sec:histology}.

\subsection{OCT system and DOCT method}
\label{sec:OCT method}
In this study, a spectral-domain OCT (SD-OCT) with an 840-nm-band light source (SLD, IW-M-D840-HP-I-CUS, Superlum, Ireland) was used.
The light source is a superluminescent diode with a center wavelength of 840 nm and a full-width-half-maximum spectral width of 100 nm.
The light is split by a 50/50 fiber coupler (F280APC 850, Thorlabs, Inc., NJ) into sample and reference arms.
In the sample arm, the light is collimated by a fiber collimator (F280APC 850, Thorlabs), deflected by a two-axis galvanometric scanner (GVS102, Thorlabs), and scans the sample through an objective (LSM02BB, Thorlabs) with a focal length of 18 mm.
The lights from the sample and reference arms are recombined by the 50/50 fiber coupler and introduced into a spectrometer (Cobra-S 800, Wasatch Photonics, NC).
The spectral interference signal is captured at a speed of 50,000 lines/s and digitized by a Camera Link frame grabber (PCIe 1433, National Instruments Corp., TX).
The axial and lateral resolutions are 3.8 \um (in tissue) and 4.9 \um, respectively.
The system sensitivity was measured as 104.9 dB with a probe power on the sample of 6.28 mW.

For volumetric DOCT imaging, a repeating raster scan protocol\cite{ElSadek2021BOE} was used.
Thirty-two repeated frames were captured at each location with a frame repeating time of 0.23 s, and 128 B-scan locations on the sample were scanned.
The \enface FOV was divided into 8 sub-fields along the slow scan direction, and each sub-field was scanned by the repeating raster protocol.
The measurement time of a single sub-field was approximately 7.35 s and the total acquisition time for the volume was 58.8 s.
Each frame and A-line comprise 512 A-lines and 1,024 pixels, respectively.
Each \enface image consists of 512 $\times$ 128 pixels for the fast-scan direction times slow-scan direction.
And hence, the \enface pixel sizes (fast-scan $\times$ slow-scan directions) are 11.7 \um $\times$ 46.9 \um for the widest FOV, 5.86 \um $\times$ 23.4 \um for the middle FOV, and 1.95 \um $\times$ 7.81 \um for the smallest FOV.
It is noteworthy that, even the OCT system has only a standard A-line rate of 50 kHz, the volumetric DOCT with \enface field size of 512 $\times$ 128 pixels could be acquired within a minute.

\subsubsection{Bulk motion correction}
\label{sec:motion correction}
The DOCT method is sensitive to bulk motions which can be caused by system and sample vibrations and environmental fluctuations during the measurement.
Such motions cause erroneously high LIV values as to be discussed in detail in Section \ref{sec:DiscussionMotionCorrection}. 
And hence, the bulk motion among the frames along the depth- and slow-scan-directions (i.e., the in-plane bulk motion) was corrected by image registration before LIV computation.
The registration is performed for the time-sequential 32 frames measured at the same location.
The 16th frame is used as a reference image and the bulk motion of each frame with respect to the reference frame is detected by sub-pixel (1/10-pixel accuracy) image registration (\mbox{skimage.registration.phase\_cross\_correlation} of scikit-image 0.17.2 in Python 3.8.5).
The detected motion is corrected by a sub-pixel shift method (scipy.ndimage.shift function of SciPy 1.5.2 using third-order spline interpolation) in the dB-scaled intensity image.

The computation time for a single frame was 0.06 s for registration and 0.06 s for the sub-pixel shift.
Because one volume comprises 4,096 frames (i.e., 32 frames times 128 B-scan locations), the total computation time is estimated to be approximately 8.2 min.

\subsubsection{Logarithmic-intensity-variance (LIV) and OCT images}
The DOCT contrast used in this study is LIV, which is defined as a variance of dB-scaled OCT signal intensity \cite{ElSadek2020BOE},
\begin{equation}
	\label{eq:LIV}
	\LIV(x,z) = \frac{1}{N}\sum_{i=0}^{N-1}\left[I_{dB}(x, z; t_i)-\langle I_{dB}(x, z; t_i) \rangle\right]^2,
\end{equation}
where $x$ and $z$ are the lateral and depth positions, respectively.
$t_i$ ($i = 1, 2, 3, \cdots$) is the sampling time of the $i$-th frame, $I_{dB}(x, z; t_i)$ is the dB-scaled OCT signal intensity, and $\langle\quad\rangle$ represents the average over time.
LIV is expected to be sensitive to the magnitude of the intratissue and/or intracellular dynamics \cite{ElSadek2021BOE, Mukherjee2021SR}.

A pseudo-color LIV image was generated by combining the OCT intensity and LIV as the brightness and hue of the image, respectively.

The OCT-intensity image was obtained by averaging all of the dB-scaled OCT images at the location.

\subsection{Histology}
\label{sec:histology}
After the DOCT measurement, the alveolar organoids were formalin fixed and paraffin embedded for hematoxylin and eosin (HE-) and elastica van Gieson (EVG-) stained histology imaging.
The EVG-stained histology highlights elastic fiber with dark blue and collagen fiber with pink.
These were sectioned with a thickness of 5 \um and mounted on a glass slide.

For HE-staining, the paraffin sections of the alveolar organoids were deparaffinized with xylene and rehydrated through a gradual concentration series, starting with 100\% to 70\% ethanol and ending with deionized water.
The sections were stained for nucleus with hematoxylin (Wako), washed with water, and quickly exposed to acidic ethanol.
Subsequently, the sections were stained with eosin targeting the cytoplasm.
Finally, these sections were dehydrated through a sequential concentration change from 70\% to 100\% ethanol.

For EVG staining, the paraffin sections of the alveolar organoids were deparaffinized and rehydrated in a descending alcohol series, and stained with an EVG staining kit (\#1.15974, Millipore, MA) according to the manufacturer’s protocol.


\subsection{Image analysis}
\label{sec:image analysis}
To quantify the image characteristics of the alveoli, the counts, area, and circularity of the alveoli were investigated.
For each time point and organoid type (i.e., normal or bleomycin model), a volume with a 3-mm $\times$ 3-mm FOV was analyzed.
An \enface image of pseudo-color LIV was extracted at approximately 100 \um below the surface from each volume.
In this analysis, the \enface plane is a flat plane which is manually tilted to become roughly parallel to the sample surface by post-processing.

The alveoli were manually segmented by an operator  (Morishita) on each \enface LIV image.
Here the alveolar region is defined as the area within the outer border of the epithelium.

Here we use the \enface rather than cross-sectional images for the image analysis for two reasons.
At first, the alveolar organoid has a flat disk-like shape.
In addition, our OCT does not visualize the full depth of the organoids as shown in Fig.\@ \ref{fig:eachDepthEnface} and Figs.\@ S1-S11 (supplementary).
And hence, it is more reasonable to use \enface images for the image-based alveolar analysis.

As discussed later in the Result and Discussion sections (Sections \ref{sec:Result} and \ref{sec:Discussion}), the alveoli appeared as cystic structures surrounded by alveolar epithelium with hyper scattering.
Some alveoli exhibited hyper scattering inclusion (denoted as filled alveoli) and others did not (non-filled alveoli).
Some alveoli had epithelia with homogeneously high LIV, while other epithelia showed tessellated patterns of high and low LIV.   
The alveoli were classified into four types based on these characteristic appearances, i.e., the combinations of non-tessellated or tessellated and filled or non-filled, as summarized in Table \ref{tab:classTable}, which also defines the abbreviations of each type.
The small alveoli without clear lumen were classified as filled alveoli.
In addition, the alveolar analysis was performed by using an \enface image for each volume.
This two-dimensional analysis was selected because we manually segmented the alveoli.
Future development of automatic 3D segmentation may enhance the reliability of the analysis.

\begin{table}
	\caption{The alveolar type classification based on the mass encapsulation (non-filled or filled) and LIV appearance (tessellated or non-tessellated).
		The table also defines the abbreviations for each type.}
	\label{tab:classTable}
	\centering
	\begin{tabular}{c|c|c|c}
		\multicolumn{2}{c|}{}&\multicolumn{2}{c}{LIV appearance of epithelium}\\
		\cline{3-4}
		\multicolumn{2}{c|}{}& Non-tessellated & Tessellated\\
		\hline
		Mass & Non-filled & T0F0 & T1F0 \\
		\cline{2-4}
		encapsulation &Filled     & T0F1 & T1F1 
	\end{tabular}
\end{table}

By using the segmentation masks, the area and circularity of each alveolus were computed for each type, where the circularity becomes 1.0 (maximum) for a perfect circle.
The counts of the alveoli were also obtained for each type.
The difference between the normal and bleomycin-model organoids was statistically analyzed.
The differences in the mean area and circularity were tested by a two-tailed Welch's t-test \cite{Welch1947}, while the differences in variance-of-area and variance-of-circularity were tested by median-based Levene’s test \cite{Brown1974JOASA}.
Alveoli overlaying the image periphery were excluded from the analysis.

Manual segmentation was performed by hand-drawing with a tablet computer and a stylus (iPad and Apple pencil, Apple, CA).
The shape analyses of the segmented alveoli were performed with custom-made software written in Python 3.8.5 with NumPy 1.19.2 and OpenCV 4.0.1 libraries.
The statistical tests were performed using Python with SciPy 1.5.2 library.

\section{Result}
\label{sec:Result}

\begin{figure}
	\centering\includegraphics{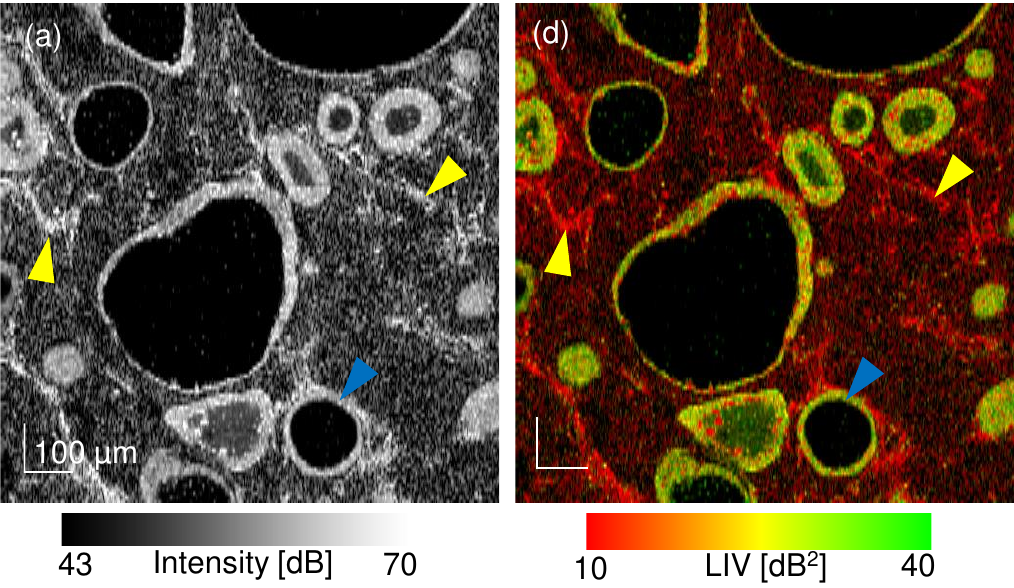}
	\caption{(a) \Enface OCT-intensity and (b) LIV images of a normal alveolar organoid measured 3 days after the organoid establishment.
		The images were extracted from approximately 100 \um below the sample surface and the FOV is 1 mm $\times$ 1 mm.
		The cystic and mesh-like structures are expected to be alveoli and fibroblasts, respectively.
	}
	\label{fig:normal organoid}
\end{figure}
An example of \enface OCT-intensity and LIV images of the normal alveolar organoid (Day +3) are shown in Fig.\@ \ref{fig:normal organoid}.
The images were extracted from approximately 100 \um below the sample surface and the FOV is 1 mm $\times$ 1 mm.
As shown in the intensity image [Fig.\@ \ref{fig:normal organoid} (a)], cystic structures (identified by a blue arrowhead) and mesh-like structures (identified by a yellow arrowhead) are visible.
As shown in the LIV image [Fig.\@ \ref{fig:normal organoid} (b)], the cystic structures are surrounded by high-LIV borders (green and red-green mixture), while the mesh-like structures have low LIV (red).
As discussed in Section \ref{sec:interpretation}, the cystic structures are believed to be alveoli, and the mesh-like structures are expected to be fibroblasts.
Therefore, they are denoted as alveoli and fibroblasts in this section.
The high-LIV border of an alveolus is believed to be alveolar epithelium.

\begin{figure}
	\centering\includegraphics{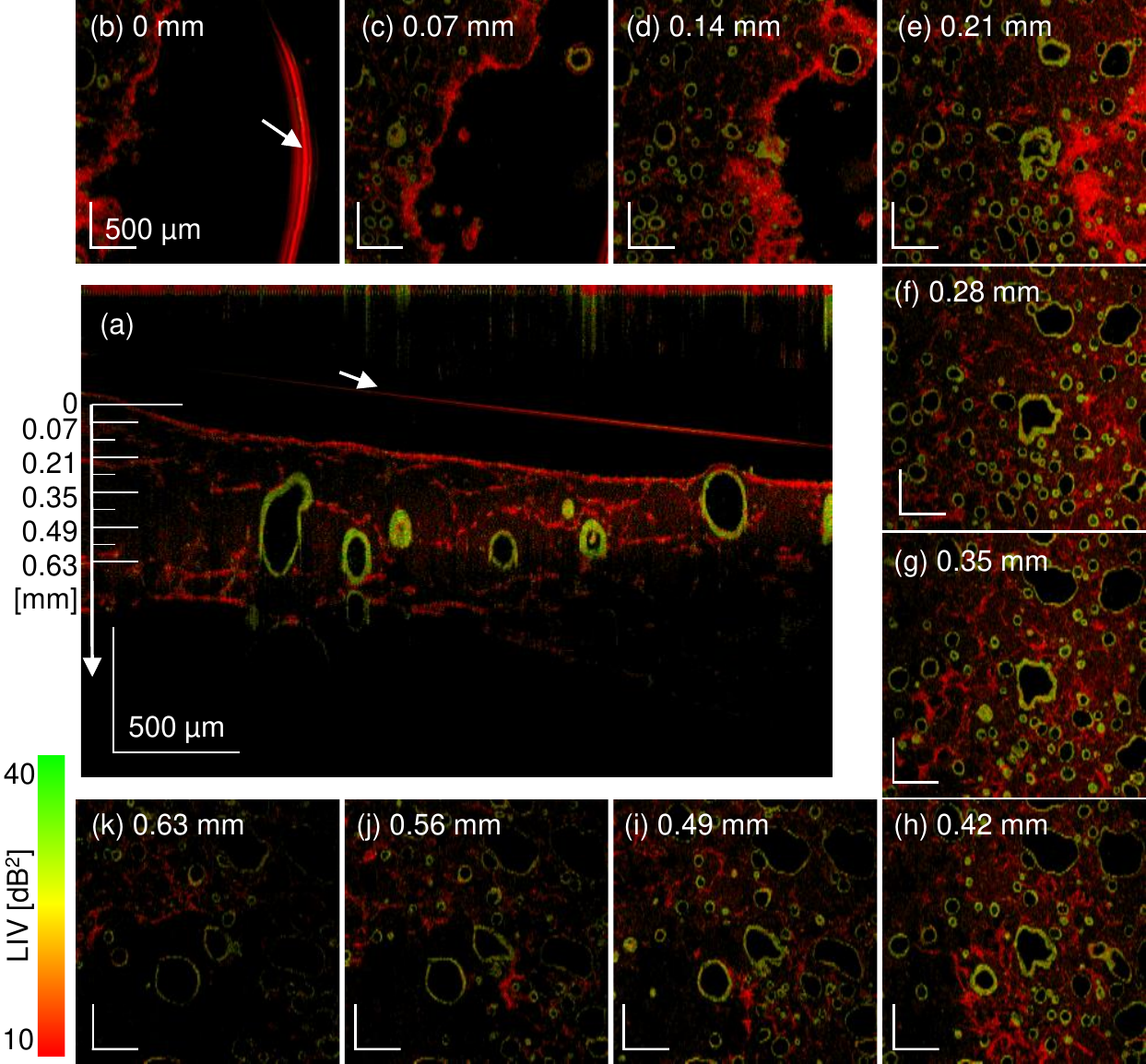}
	\caption{
		The LIV \enface images of the Day +3 normal organoid at several depths (square panes, FOV is 3 mm $\times$ 3 mm).
		The images were extracted at a depth of every 0.07 mm, and the depth positions are indicated in the cross-sectional LIV image at the center.
		The alveoli with high LIV (green or green-red mixture) borders, which are possibly alveolar epithelium, and the fibroblasts are uniformly distributed in 3D.
		The red arc at the 0-mm depth (arrow) and the red line in the cross-sectional image (arrow) are caused by the surface reflection of the culture medium.}
	\label{fig:eachDepthEnface}
\end{figure}
The alveoli and fibroblasts are observed throughout the whole image depth as shown in Fig.~\ref{fig:eachDepthEnface}.
The sample is the Day +3 normal organoid, and the \enface FOV is 3 mm $\times$ 3 mm.
The \enface LIV images (square panes) every 0.07 mm from the top to the bottom are shown, and the depth positions of these images are indicated in the cross-sectional LIV image at the center.
Depth-independent distribution of the alveoli and fibroblasts was observed in all samples.
All the samples are visualized in the same manner in Supplementary Figs.\@ S1-S11.

\begin{figure}
	\centering\includegraphics{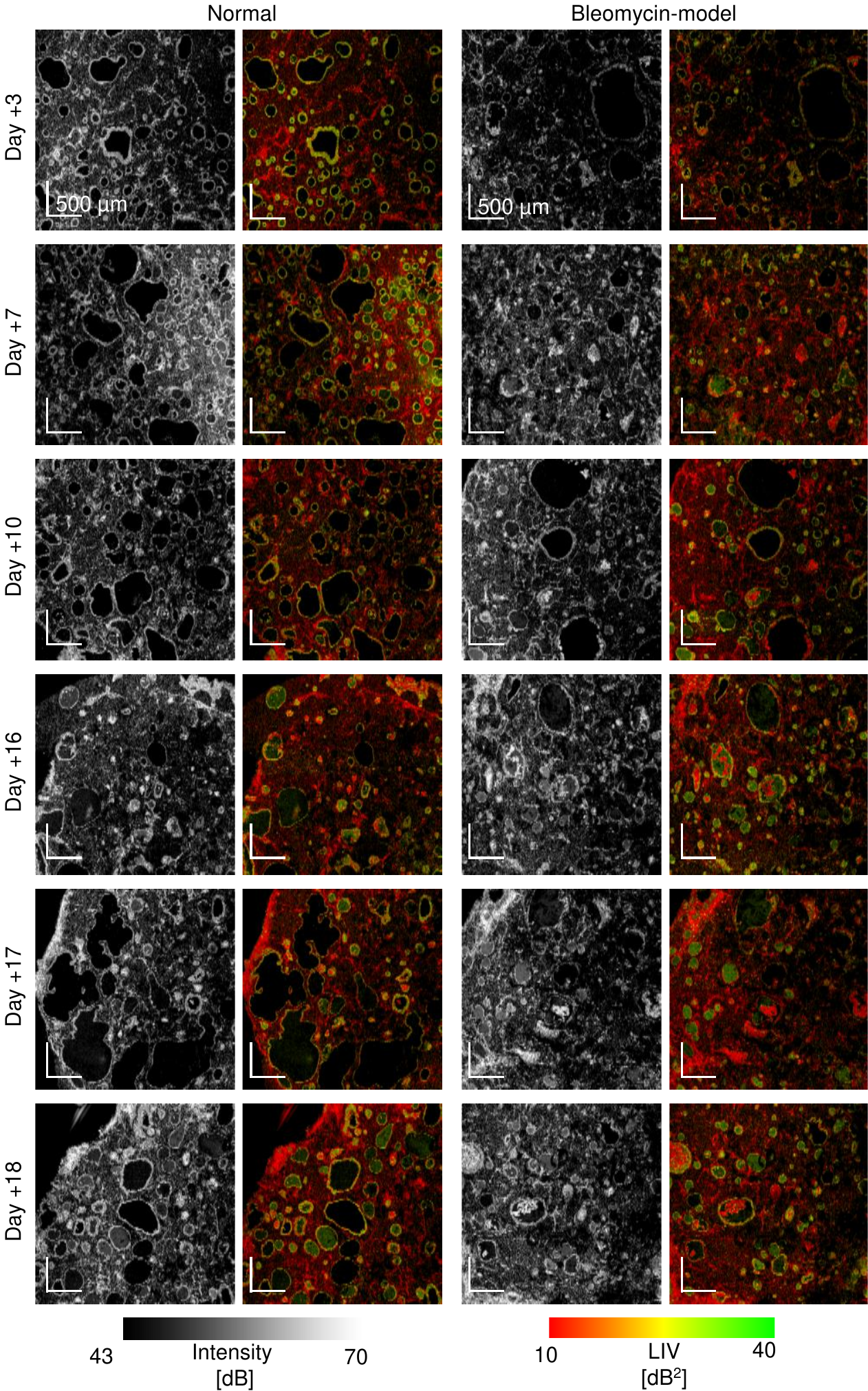}
	\caption{
		The \enface OCT-intensity and LIV images of normal and bleomycin-model alveolar organoids at Days +3, +7, +10, +16, +17, and +18.
		The images were extracted approximately 100-\um below the sample surface.
		The FOV is 3 mm $\times$ 3 mm.}
	\label{fig:enface_allTP}
\end{figure}
\begin{figure}
	\centering\includegraphics{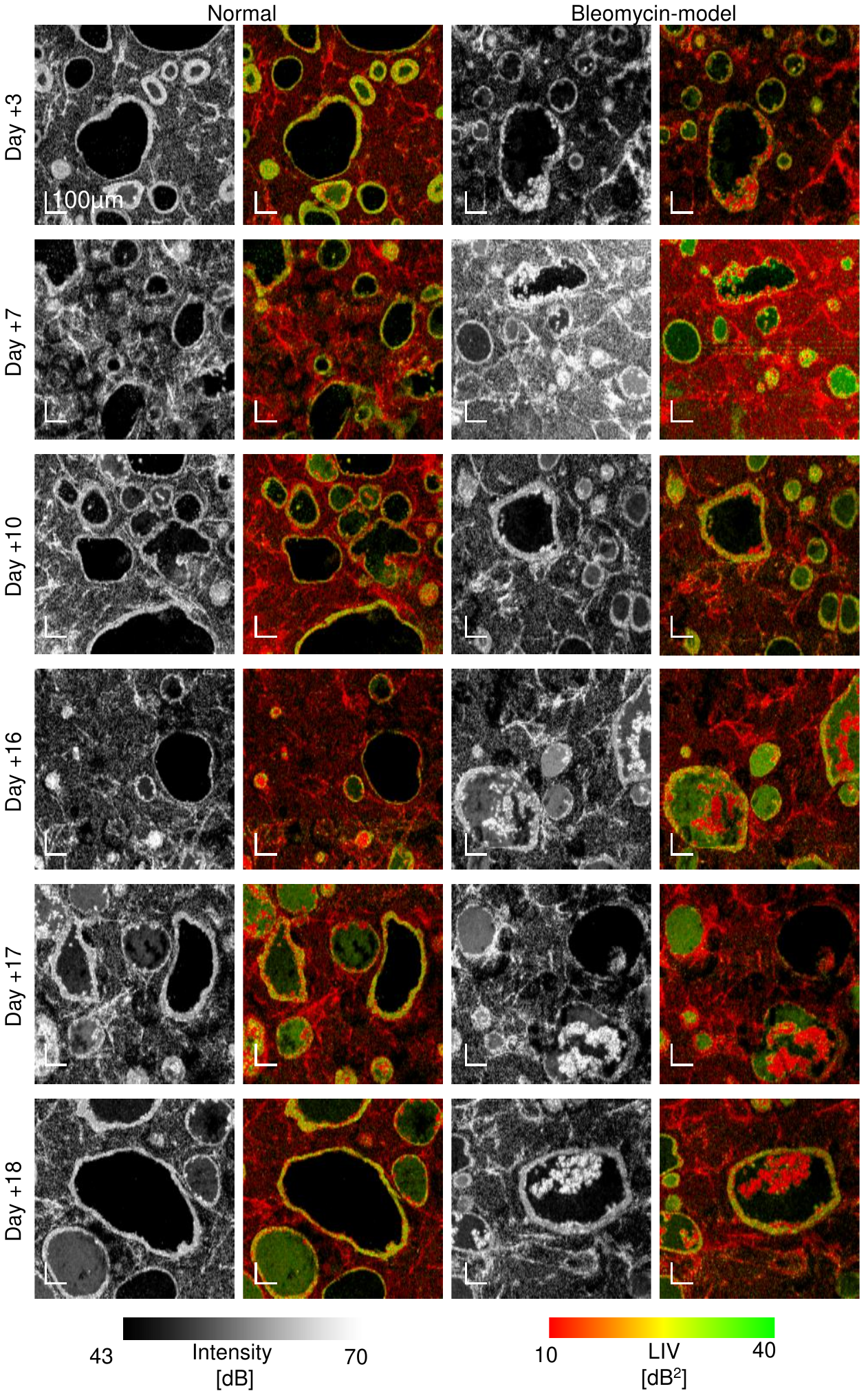}
	\caption{
		The \enface OCT-intensity and LIV images at all time points. 
		The samples are the same with Fig.\@ \ref{fig:enface_allTP} but measured with a smaller FOV of 1-mm $\times$ 1-mm.
		Due to the smaller pixel separation (i.e. higher pixel density) than that of Fig.\@ \ref{fig:enface_allTP}, finer structures are visible.
		The images were extracted approximately 100-\um below the sample surface.
		The normal model at Day +3 is the same image with Fig.\@ \@\ref{fig:normal organoid}.
	}
	\label{fig:enface_allTP_1mm}
\end{figure}
The \enface OCT-intensity and LIV images of the normal and bleomycin-model organoids at all time points (Days +3, +7, +10, +16, +17, and +18) are shown in Figs.\@ \ref{fig:enface_allTP} and \ref{fig:enface_allTP_1mm}.
These figures show the same samples.
The FOV of Fig.\@ \ref{fig:enface_allTP} is 3 mm $\times$ 3 mm, while that of Fig.\@ \ref{fig:enface_allTP_1mm} is 1 mm $\times$ 1 mm.
Since Fig.\@ \ref{fig:enface_allTP_1mm} has higher magnification (i.e., smaller pixel separation), it reveals finer structures than Fig.\@ \ref{fig:enface_allTP}.
It should be noted that the images are not a real longitudinal imaging result but a pseudo-time-sequence.
Namely, although the samples are cultured in the same protocol, they are not the same.
For all samples, the cystic structures with high-LIV borders (alveoli with high-LIV alveolar epithelium) and low-LIV mesh-like structures (fibroblasts) were observed.
The size and shape of the alveoli significantly varied.
The sizes of the large and small alveoli showed around ten-fold difference.
Additionally, the shape of the alveoli also widely varied from highly circular to significantly deformed.
This circularity ranged from approximately 0.9 to 0.2, as discussed later.

\begin{figure}
	\centering\includegraphics{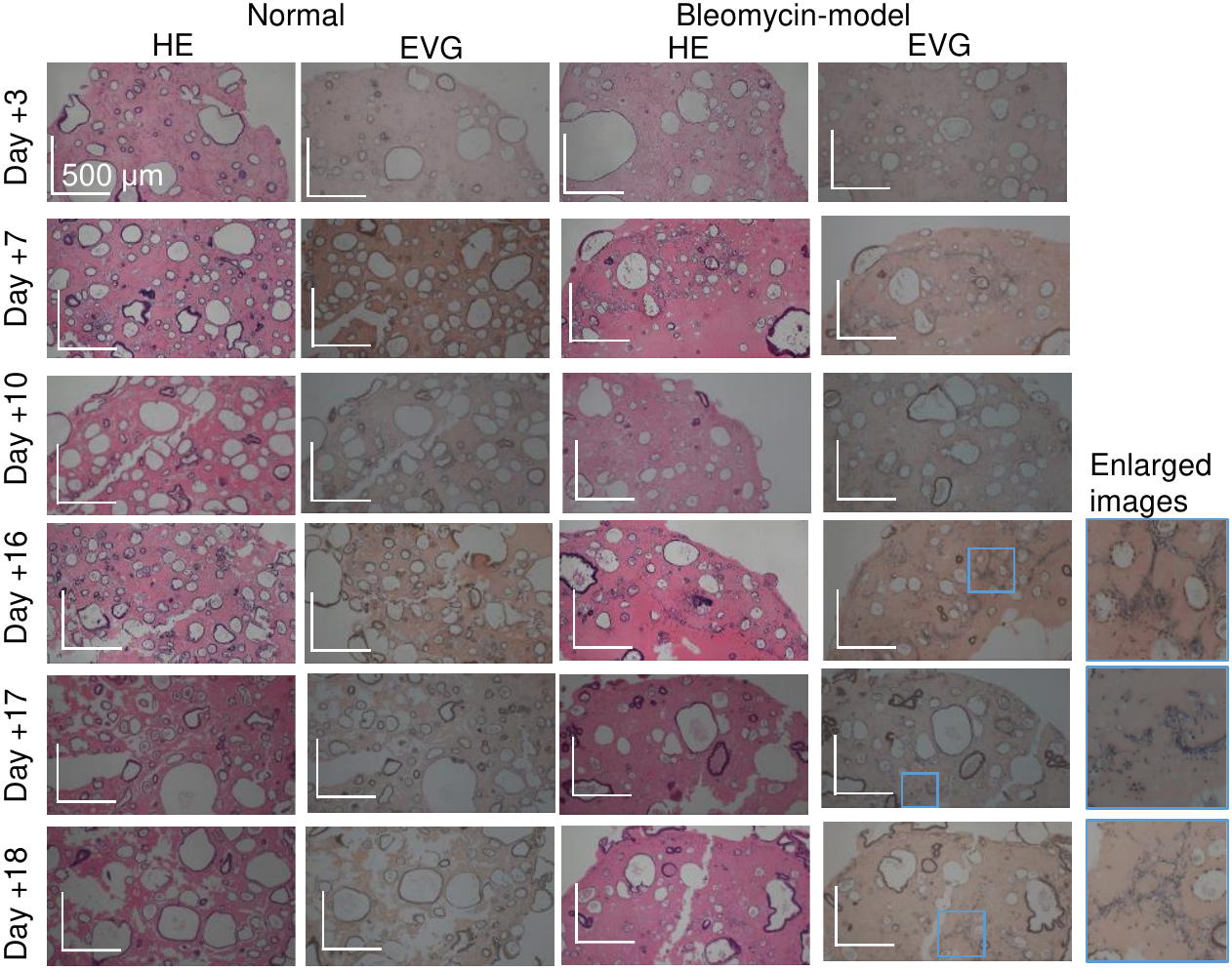}
	\caption{
		The HE- and EVG-stained histological micrographs obtained from the identical samples to the OCT and LIV images of Fig.\@ \ref{fig:enface_allTP}.
		Appearances similar to the corresponding OCT images were observed in all samples.
		In the EVG images, elastin appeared as dark blue.
		In the enlarged bleomycin-model organoids (the most right column), the fibroblasts appeared as dark blue in the EVG image. 
	}
	\label{fig:histology}
\end{figure}
The OCT appearances correlated well with the HE- and EVG-stained histological micrographs shown in Fig.\@ \ref{fig:histology}.
Here, the histological micrographs were obtained from the identical samples to the OCT and LIV images of Fig.\@ \ref{fig:enface_allTP}.
The similar shape and distribution of the cystic structures to the OCT and LIV images were observed in all histological micrographs.
Elastin appeared as dark blue in the EVG images.
In the late time points (Days +16, +17, and +18), the fibroblasts show the dark blue appearance (see the most right column).

\begin{figure}
	\centering\includegraphics{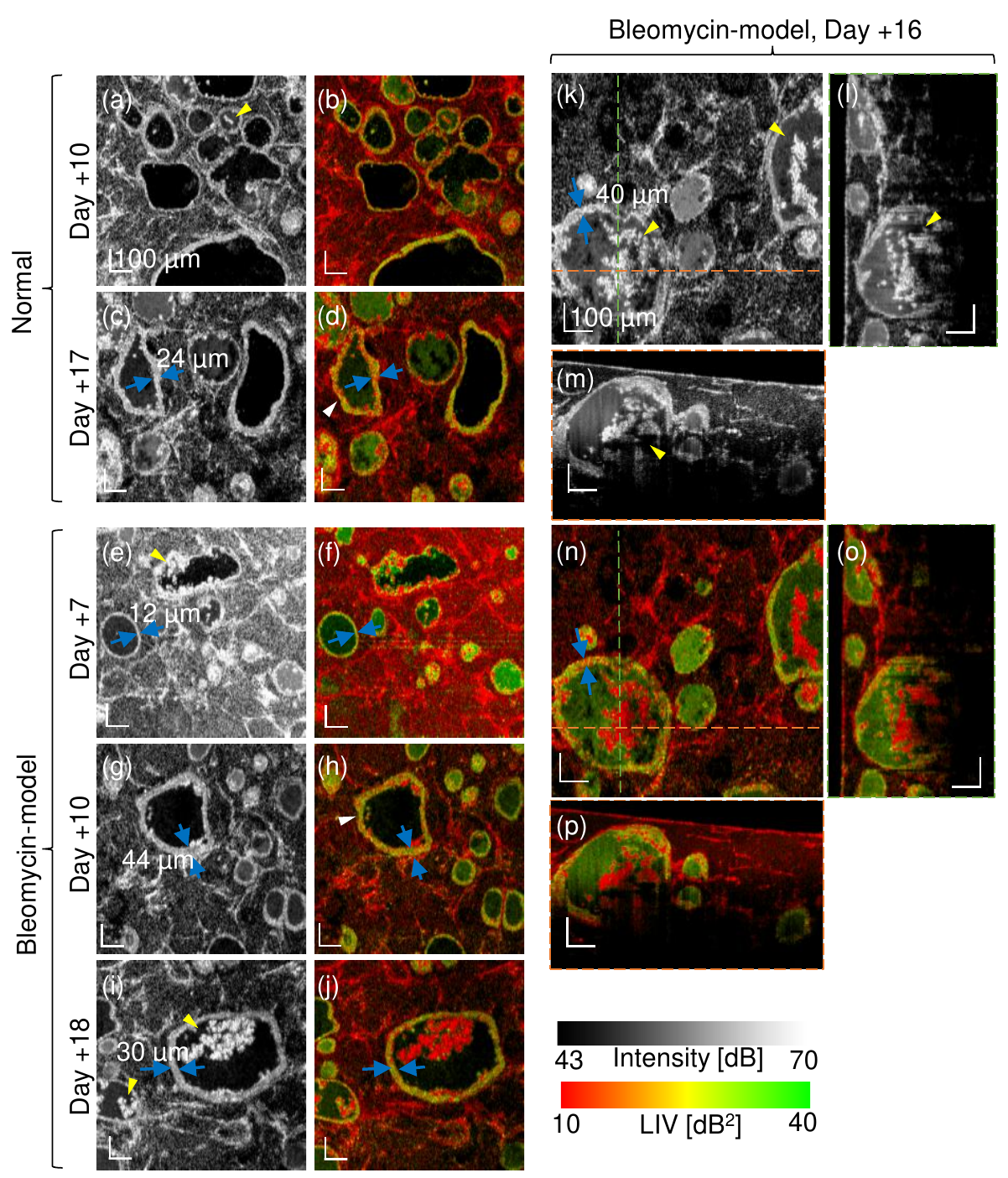}
	\caption{Selected OCT-intensity and LIV images showing several types of alveoli.
		The samples are the Days +10 and +17 normal model and the Days +7, +10, +16 and +18 bleomycin model.
		The FOV of the \enface image is 1 mm × 1 mm.
		(l), (m), (o) and (p) are the cross-sections at the green and orange lines in the corresponding \enface image (k) and (n).
		The tessellated dynamic appearance of the alveolar epithelium is denoted by the white arrows, and the ragged alveoli that encapsulate the hyper-scattering mass are indicated by yellow arrows.
		The \enface images are a subset of Fig.\@ \ref{fig:enface_allTP_1mm}.}
	\label{fig:alveoli}
\end{figure}
Some alveolar organoids showed cystic structures with a thickened and ragged border (possibly thickened and ragged alveolar epithelium), as shown in Fig.\@ \ref{fig:alveoli}.
Here the FOV of the \enface images is 1 mm $\times$ 1 mm.
The images (l), (m), (o), and (p) are the cross-sectional images at the green and orange dotted lines in the corresponding \enface images [(k) and (n)].
The cystic structures with a ragged border encapsulate the hyper-scattering mass, as denoted by the yellow arrowheads in the OCT-intensity images, and/or show the tessellated high-and-low LIV appearance in its border (epithelium) as denoted by the white arrowheads in the LIV images.

To further analyze the shape of the alveoli (cystic structures) and the LIV appearance of its border (alveolar epithelium), we computed the count, circularity, and area of the alveoli.
The alveoli were categorized into four types based on the hyper-scattering-mass encapsulation and the LIV appearance of the epithelium, as summarized in Table \ref{tab:classTable}.

\begin{figure}
	\centering\includegraphics{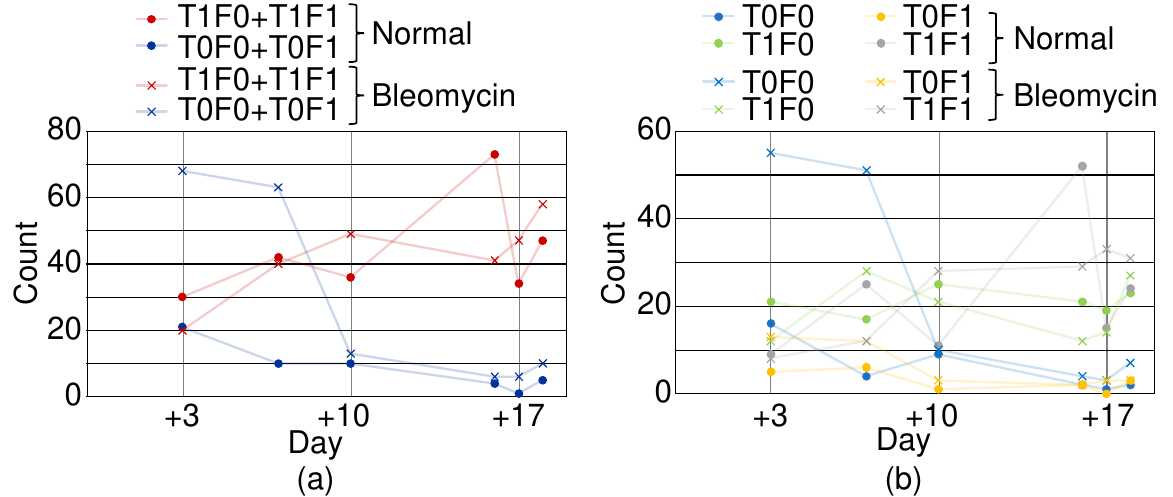}
	\caption{
		(a) Counts of the tessellated and non-tessellated alveoli.
		Here the counts are the summations of filled and non-filled alveoli.
		(b) Count of all four types of alveoli. 
		The counts in the normal and bleomycin models are plotted with dots and cross marks, respectively.
		The counts were obtained using \enface images with a FOV of 3 mm $\times$ 3 mm as described in Section \ref{sec:image analysis}.
		As shown in (a), the tessellated alveoli (red) show an increasing trend over time for the both normal and bleomycin-model organoids.
		In contrast, the non-tessellated alveoli (blue) show a decreasing trend over time for both models.
	}
	\label{fig:count}
\end{figure}
Figure \ref{fig:count} shows the counts of alveoli for each type.
As shown in Fig.\@ \ref{fig:count}(a), the combined counts of the two types of tessellated alveoli (red, T1F0 + T1F1) increase over time for both normal and bleomycin-model organoids, while the combined counts of the non-tessellated alveoli (blue, T0F0 + T0F1) decrease over time for both models.
By independently assess th four alveolar types, remarkable normal-to-bleomycin difference was found at the early time points (Days +3 and +7) for the non-tessellated and non-filled alveoli (blue, T0F0) as shown in Fig.\@ \ref{fig:count}(b).
Namely, the alveoli of the bleomycin model showed evidently greater counts than those of the normal organoids.

\begin{figure}
	\centering\includegraphics{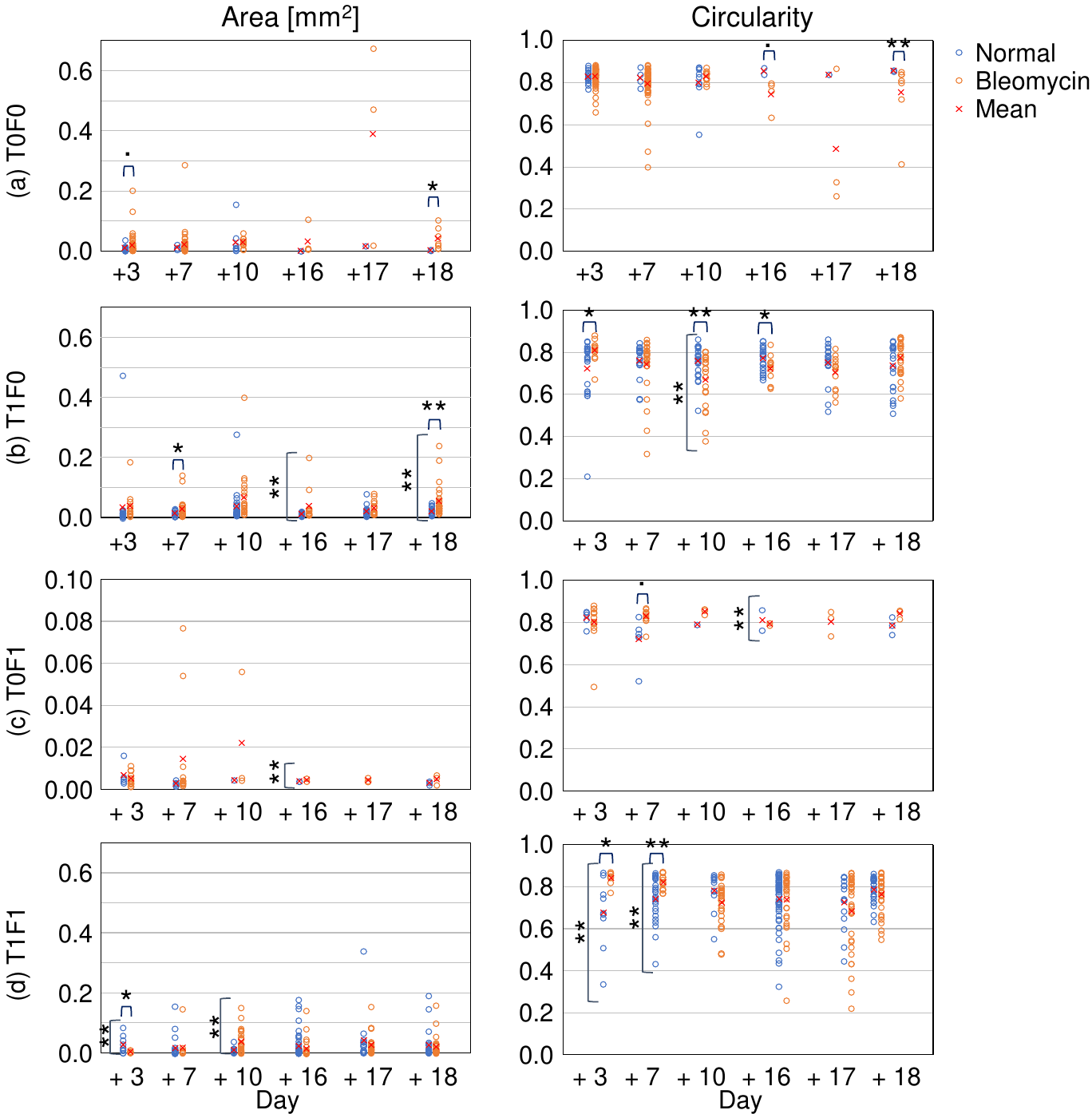}
	\caption{
		Area and circularity of the four types of alveoli.
		Each circle corresponds to each organoid, and the mean values are plotted with cross marks.
		Significant differences of mean and variance between the normal and bleomycin-model organoids are marked on the top and side of the plots, respectively.
		The dot ($\cdot$), $\ast$, and $\ast\ast$ indicate P $<$ 0.1, 0.05, and 0.01, respectively.}
	\label{fig:area-circularity}
\end{figure}
Figure \ref{fig:area-circularity} shows the area and circularity of the alveoli of each type.
Significant differences in mean and variance between the normal and bleomycin models are marked on the top and side of the plots, respectively.
Specifically, the dot ($\cdot$), $\ast$, and $\ast\ast$ indicate P $<$ 0.1, 0.05, and 0.01, respectively.

The areas are widely diverse from approximately $4\times 10^{-6}$ to $0.7$ mm$^2$, and the circularity are also widely varied from approximately 0.2 to 0.9.
For T0F0 at the early time points (Days +3 and +7) [Fig.\@ \ref{fig:area-circularity}(a)], the normal organoids (blue) showed smaller and more circular alveoli than the bleomycin model (red).
On the other hand, for T1F1 [Fig.\@ \ref{fig:area-circularity}(d)], the normal organoids (blue) showed larger (at Day +3) and less circular (at Days +3 and +7) alveoli than the bleomycin-model organoids (red).
At the late time points (Days +16, +17, and +18) of T1F0 [Fig.\@ \ref{fig:area-circularity}(b)], the bleomycin-model organoids (red) showed larger means and variance in the areas than the normal organoids (blue).
Note that we did not discuss the cases with alveolar counts of 7 or less, even if significant differences were observed.

It should be noted that, for the analyses presented in Figs.\@ \ref{fig:count} and \ref{fig:area-circularity}, we used only one organoid for each combination of the models and time-points.
And hence, the results cannot be well generalized.
It might be a future work to account for this issue by increasing the number of samples.

\section{Discussion}
\label{sec:Discussion}

\subsection{Alveolar organoids in OCT-intensity and LIV images}
\label{sec:interpretation}

\subsubsection{Alveoli and fibroblasts}
The cystic structures and mesh-like structures were observed in the OCT-intensity and LIV images.
By considering the culture process of the alveolar organoids, the alveolar organoids must contain only two kinds of structure-building cells, i.e., the hiPSC-derived alveolar epithelial cells and one type of mesenchymal cell, i.e., fibroblasts.
The alveolar epithelial cells form cysts that mimic alveoli in the alveolar organoids \cite{Yamamoto2017NM}.
Therefore, we can conclude that the cystic structure observed in the OCT-intensity and LIV images is alveolus and its hyper-scattering border is composed of alveolar epithelial cells.
In addition, the mesh-like structures are expected to be constructed by the other structure building cells, i.e., fibroblasts.

\begin{figure}
	\centering\includegraphics{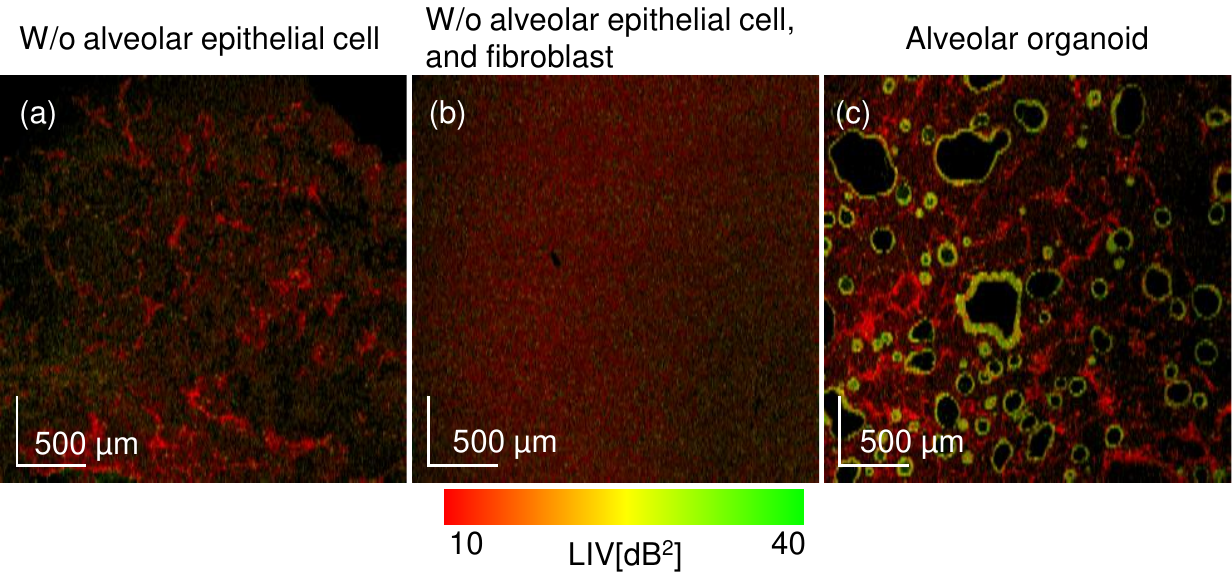}
	\caption{
		The comparison of \enface LIV images of a sample comprising only fibroblasts and Matrigel (a), pure Matrigel (b), and the Day +3 normal organoid (c).
		(c) was reprinted from Fig.\@ \ref{fig:enface_allTP} for reference.
		By comparing these images, it is evident that the mesh-like structures are fibroblasts.
		The FOV of \enface image is 3 mm $\times$ 3 mm.}
	\label{fig:matrigel fibroblast}
\end{figure}
These interpretations of the cystic and mesh-like structures were directly validated by measuring two more samples.
One is a sample comprising fibroblasts and Matrigel (i.e., without alveolar epithelial cells).
The other is a pure Matrigel.
The results are summarized in Fig.\@ \ref{fig:matrigel fibroblast}.
Figure \ref{fig:matrigel fibroblast}(a) shows the \enface LIV image of the sample with fibroblasts and Matrigel, while Fig.\@ \ref{fig:matrigel fibroblast}(b) is that of the pure Matrigel sample.
Figure \ref{fig:matrigel fibroblast}(c) shows a normal organoid at Day +3, which is reprinted from Fig.\@ \ref{fig:enface_allTP} for comparison.
By comparing these images, it is evident that the mesh-like structures are fibroblasts and not alveolar epithelium or Matrigel.

In addition, in the late time points of the bleomycin model (Day +16, +17, and +18), the elastin structures (dark blue) in the EVG-stained images [Fig.\@ \ref{fig:histology}, the most right column] showed pattern similarity with the mesh-like structures in the OCT-intensity and LIV images.
Because elastin is generated in the fibroblast, this structural similarity further supports our interpretation of the mesh-like structure as fibroblasts.

\subsubsection{Thickened alveolar epithelium}
Thickening of the alveolar epithelium was observed in the OCT-intensity images.
The thicknesses of the alveolar epithelia indicated with blue arrows in Figs.\@ \ref{fig:alveoli}(c), (e), (g), (l), and (k) were manually measured.
The thicknesses of the thickened epithelia [Figs.\@ \ref{fig:alveoli}(c), (g), (l), and (k)] range from 24 \um to 44 \um, which is twice to four times thicker than that of the non-thickened epithelium [12 \um, Fig.\@ \ref{fig:alveoli}(e)].

It is known that abnormal repairing and remodeling cause the thickening of alveolar epithelium \cite{Carraro2020AJ, Desai2014Nature}.
Bronchiolization is one type of abnormal repair and remodeling \cite{Murthy2022Nature}, where the alveolar epithelium differentiates into the bronchial epithelium \cite{Plantier2011Thorax}.
Bleomycin can induce abnormal repair including bronchiolization and differentiation to other types of alveolar epithelial cells \cite{Usuki1995PI,Suezawa2021SCR}.
Therefore, the thickened epithelium in the bleomycin-model organoid may indicate abnormal repair and remodeling such as bronchiolization.

In addition to the morphological change, tessellated patterns of high and low LIV were observed at the thickened epithelium in Fig.\@ \ref{fig:alveoli}.
Such alveolar epithelium is suspected to have undergone abnormal repair and remodeling, and hence is a mixture of the alveolar epithelial cells and other cells, e.g., bronchial epithelial cells\cite{Murthy2022Nature, Carraro2020AJ}.
Therefore, it is reasonable to consider that the abnormally remodeled cells and the alveolar epithelial cell might have different intracellular dynamics.
Further investigation by introducing more cell-type specific methods, such as immunohistochemistry, is an important future study.

The alveoli with thin epithelium, especially at the early time points, showed high LIV (Figs.\@ \ref{fig:normal organoid}, \ref{fig:eachDepthEnface}, \ref{fig:enface_allTP}, and \ref{fig:enface_allTP_1mm}).
It suggests that sound alveolar epithelial cells may exhibit high LIV.
Similarly, the high LIV regions in the tessellated epithelia may correspond to the alveolar epithelial cells.
And in contrary, the low LIV region may correspond to abnormally remodeled cells.

The LIV tessellation of the alveolar epithelium was more frequently observed in the bleomycin-model organoids and the late-time-point-normal organoids as suggested by Fig.\@ \ref{fig:count}(a).
We empirically know that the differentiation and dedifferentiation occur in the long-time-cultured alveolar epithelium.
And hence, we guess that the tessellated pattern in the normal model indicates differentiation and dedifferentiation due to the long-time culture.

In this section, we have discussed only the tessellated thick epithelium.
Further investigation and interpretation of the tessellated thin epithelium might be an important future work.

\subsubsection{Hyper-scattering mass encapsulated in alveolus}
\begin{figure}
	\centering\includegraphics{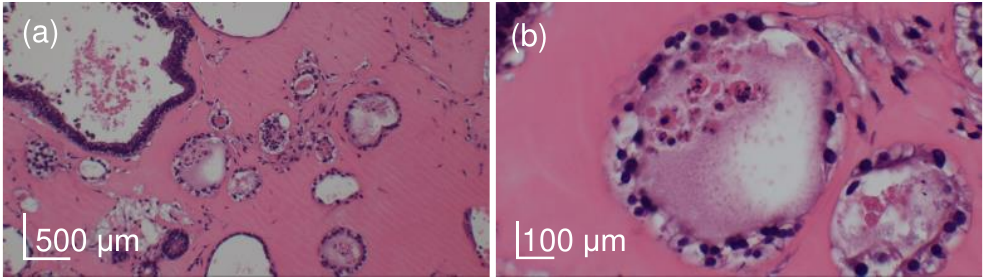}
	\caption{HE stained images of the alveoli that encapsulate some masses.
		The sample is the bleomycin model of Day +16, and the (a) and (b) are obtained from the same sample.
		Cell nuclei are observed in the mass inside of the alveoli.}
	\label{fig:HE mass}
\end{figure}
The hyper-scattering mass was observed inside some alveoli at the late time points, as shown in Fig.\@ \ref{fig:alveoli}(i) and (k) (yellow arrows).
Most alveoli encapsulating the mass have a thickened epithelium.
Similar alveoli in the HE-stained images show nuclei in the alveoli, as shown in Fig.\@ \ref{fig:HE mass}. It suggests that the hyper-scattering mass is derived from cells.
And we specifically suspect that the hyper-scattering masses are shed alveolar epithelial cells.
Although we have shown that alveolar epithelial cells exhibit high LIV, these hyper-scattering masses exhibit low LIV, as shown in Fig.\@ \ref{fig:alveoli}(j) and (n).
Therefore, the hyper-scattering masses are shed abnormally remodeled or abnormally differentiated/dedifferentiated epithelial cells.

\subsubsection{Open issue}
The alveolar counts [Fig.\@ \ref{fig:count}(b)] showed the remarkable normal-to-bleomycin difference for non-tessellated and non-filled (T0F0) alveoli only at the early time points (Days +3 and +7).
In general, the bleomycin model is formed by 3- to 6-day application of bleomycin \cite{Suezawa2021SCR}.
And hence, it is reasonable to see the difference at the early time points in our study.
On the other hand, the behavior of the bleomycin model at late time points was not well investigated yet, and it might be a future study.

\subsection{Bulk motion correction for LIV computation}
\label{sec:DiscussionMotionCorrection}
\begin{figure}
	\centering\includegraphics{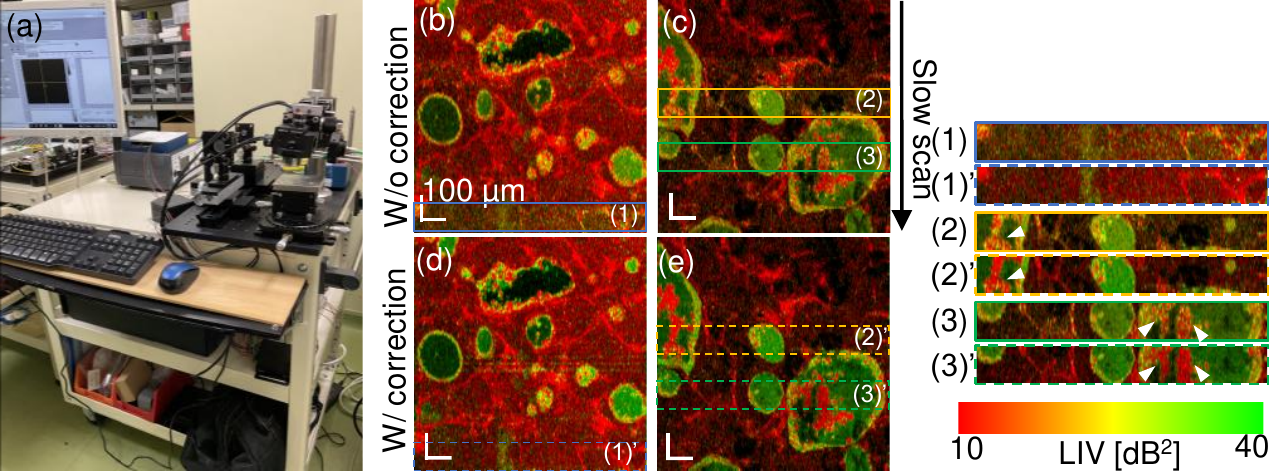}
	\caption{
		The SD-OCT system used in this study (a) and \enface LIV images without (b, c) and with (d, e)  motion correction described in Section \ref{sec:motion correction}.
		The motion artifacts are found in some sub-fields in (b) and (c) as indicated by the color boxes (blue, yellow, and green).
		These artifacts are not observed in the images with motion correction, see dashed color boxes (d, e).
		The effects of the motion correction are more readily observed in the magnified images at the right (significant points are indicated by arrowheads).
	}
	\label{fig:motionCorrection}
\end{figure}

The OCT system used in this study is an SD-OCT device built on a carrying cart, as shown in Fig.\@ \ref{fig:motionCorrection}(a).
And hence, it is less stable than our previous DOCT system built on a rigid optical bench \cite{ElSadek2020BOE, ElSadek2021BOE, Mukherjee2021SR}. 
This portable design made the measurement vulnerable to the environmental vibration and necessitates the software-based motion correction described in Section \ref{sec:motion correction}.

The effectiveness of the software-based motion correction is demonstrated in Fig.\@ \ref{fig:motionCorrection}.
In an \enface LIV image without motion correction (b), one of the sub-fields (blue box) showed erroneously high LIV values (diffusive yellow-green appearance).
This diffusive high-LIV artifact was removed by the motion correction as shown in (d) (dashed blue box).
This sub-field was magnified and compared by placing one above the other at the right side of the figure.
Another \enface LIV image without motion correction (c) exhibits two sub-fields with the LIV artifacts (yellow and green boxes).
In these fields of the non-motion corrected images, some hyper-scattering masses exhibit tessellated LIV appearance (see arrowheads in the magnified images at the right of the figure).
As shown in (e), these tessellated LIV becomes low LIV (red) after the motion correction (yellow and green dashed boxed), and hence they are artifacts.

The software motion correction is crucial for enabling LIV imaging with the portable implementation.
In addition, it should be noted that, in our motion correction, the bulk motion along the slow scan axis was not corrected.
Although this lack of correction did not affect the present LIV imaging, it potentially disturbs the imaging if the bulk motion is large.
So, further development of motion correction methods is also important to make the portable implementation more robust.

\subsection{Future perspective}
In the current study, we manually measured the thickness of the alveolar epithelium for only a small number of alveoli.
This limited number of alveoli is mainly because of the lack of automatic and accurate segmentation methods for the inner and outer edges of the epithelium.
In the future, such a segmentation method enables quantitative evaluation of the epithelial thickening and can make the epithelial thickness as a biomarker for the alveolar organoid assessment.

\section{Conclusion}
In this paper, we demonstrated label-free intratissue and intracellular dynamics imaging of alveolar organoids by using DOCT.
The alveolar epithelium showed high or tessellated high-and-low LIV appearances.
This tessellated appearance can be considered to indicate abnormal repairing and remodeling.
And hence, LIV imaging can be used for assessing abnormal repairing and remodeling, such as bronchiolization, of the alveolar organoids.
In the future, DOCT can be a useful tool for alveolar-organoid based medical, biological, and pharmaceutical research.

\section*{Acknowledgments}
The authors greatly appreciate fruitful technical discussions with Atsushi Kubota (Sky Technology).
The early contribution in the SD-OCT system development by Daisuke Oida (University of Tsukuba, and currently with Think-Lands Co., Ltd.) are gratefully acknowledged.
Rion Morishita and Toshio Suzuki are equal contributors to this work and designated as co-first authors.

\section*{Funding}
Core Research for Evolutional Science and Technology (JPMJCR2105);
Japan Society for the Promotion of Science (18H01893, 21H01836, 22K04962, 22F22355);
Austrian Science Fund (Schr\"odinger grand J4460);
Japan Science and Technology Agency (JPMJMI18G8).

\section*{Disclosures}
Morishita, Mukherjee, Abd El-Sadek, Lim, Makita, Tomita, Yasuno: Yokogawa Electric Corp. (F), Sky Technology (F), Nikon (F), Kao Corp. (F), Topcon (F), Tomey Corp (F).
Suzuki: Chugai Pharmaceutical Co, Ltd (R), Nippon Boehringer Ingelheim (R), Sysmex Corporation (R).
Lichtenegger: None.
Yamamoto: HiLung Inc. (I, P).
Nagamoto: HiLung Inc. (I).

\section*{Data availability}
The data that support the findings of this study are available from the corresponding author upon reasonable request.

\section*{Supplemental document}
See Supplement 1 for supporting content.

\bibliography{2022_alveolarOrganoid.bib}

\pagebreak
\title{Supplementary Material: Supplement 1}

\begin{abstract}
	In this supplementary material, we provide eleven figures, Figs.\@ \ref{fig:SupDay3Bleomycin}--\ref{fig:SupDay18Bleomycin}.
	These figures show the LIV images of the alveolar organoids at several depth, and supplement Fig.\@ 3 of the full-length manuscript.
	These samples are normal and bleomycin-model alveolar organoids at Days +3, +7, +10, +16, +17,and +18, while Fig.\@ 3 of the full-length manuscript presents the normal organoid at Day +3.
	The field of view of the \enface images (square panes) is 3 mm $\times$ 3 mm.
	The depth-independent distribution of the alveoli and fibroblasts were found in all samples.
\end{abstract}

\setcounter{figure}{0}
\renewcommand\thefigure{S\arabic{figure}}    
\begin{figure}[h]
	\centering\includegraphics{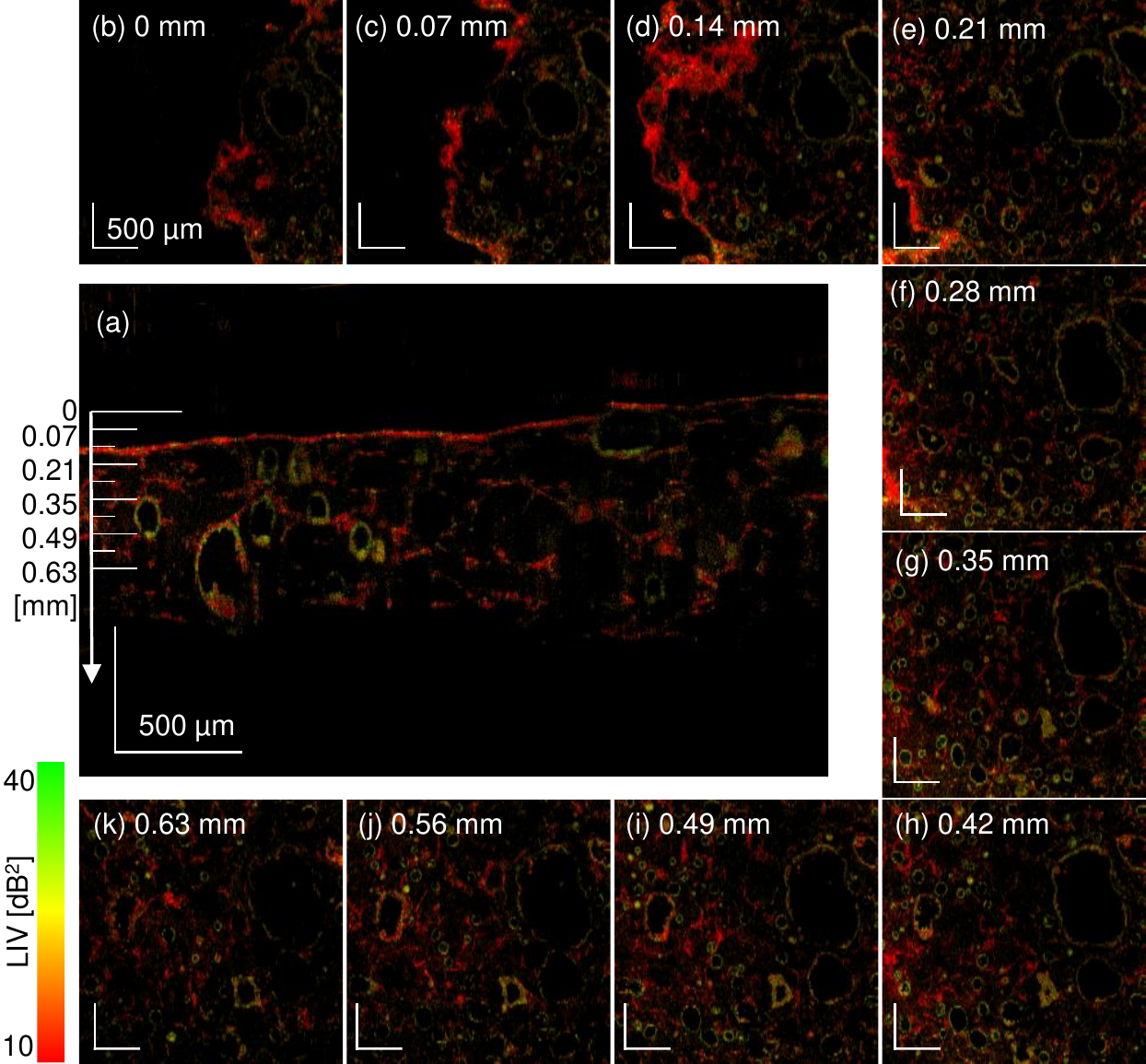}
	\caption{%
		The LIV \enface images of Day +3 bleomycin-model organoid at several depths (square panes, FOV is 3 mm $\times$ 3 mm).
		The images were extracted at the depth of every 0.07 mm, and the depth positions are indicated in the cross-sectional LIV image at the center.
		The alveoli with high-LIV (green or green-red mixture) borders, which are possibly alveolar epithelium, and the fibroblasts are uniformly distributed in 3D.}
	\label{fig:SupDay3Bleomycin}
\end{figure}

\begin{figure}
	\centering\includegraphics{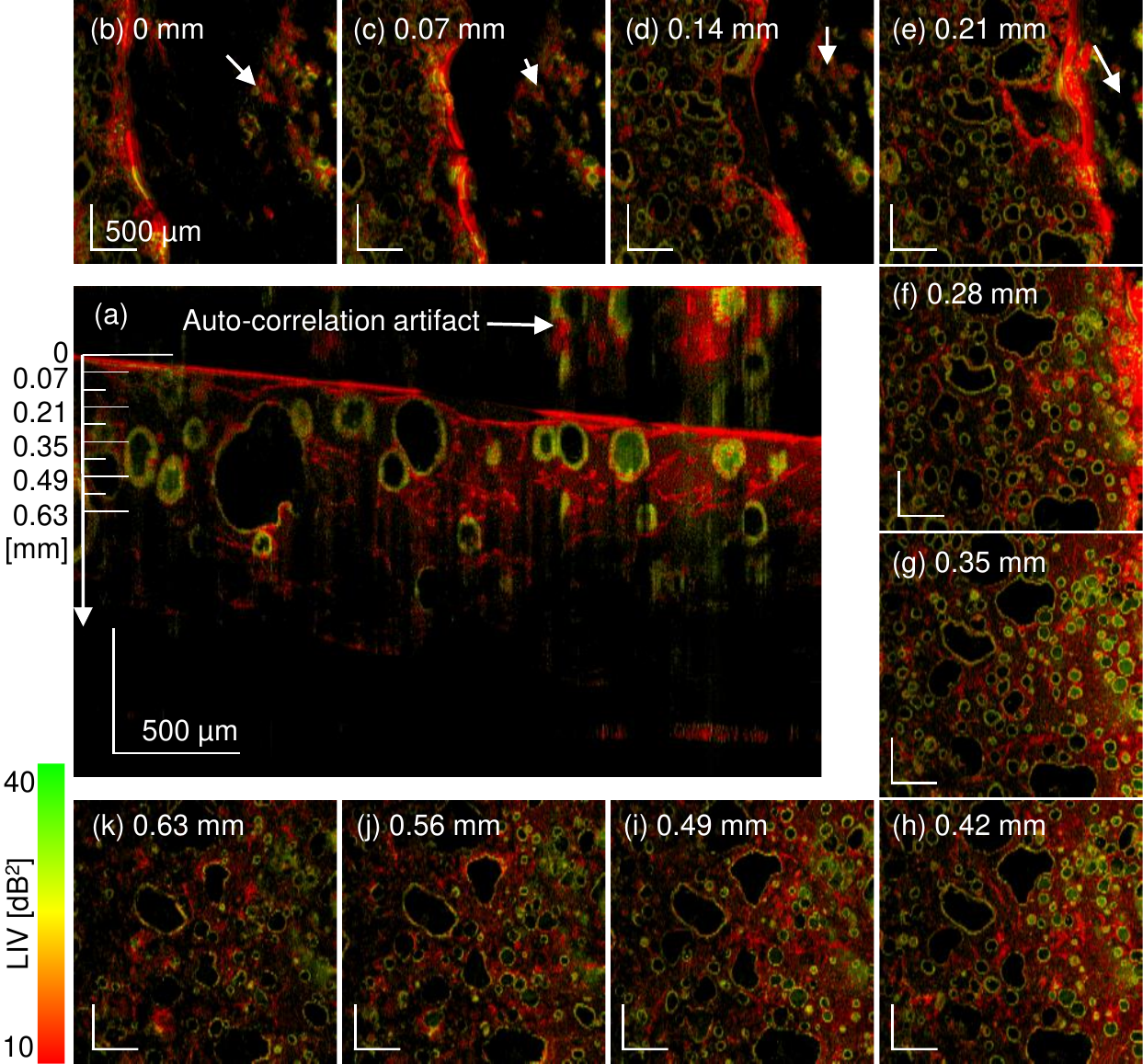}
	\caption{%
		The case of Day +7 normal organoid.
		The appearances indicated by white arrows are auto-correlation artifacts.
		Here after, all figures are presented in the same manner with Fig.\@ \ref{fig:SupDay3Bleomycin}.
	}
	\label{fig:SupDay7Normal}
\end{figure}

\begin{figure}
	\centering\includegraphics{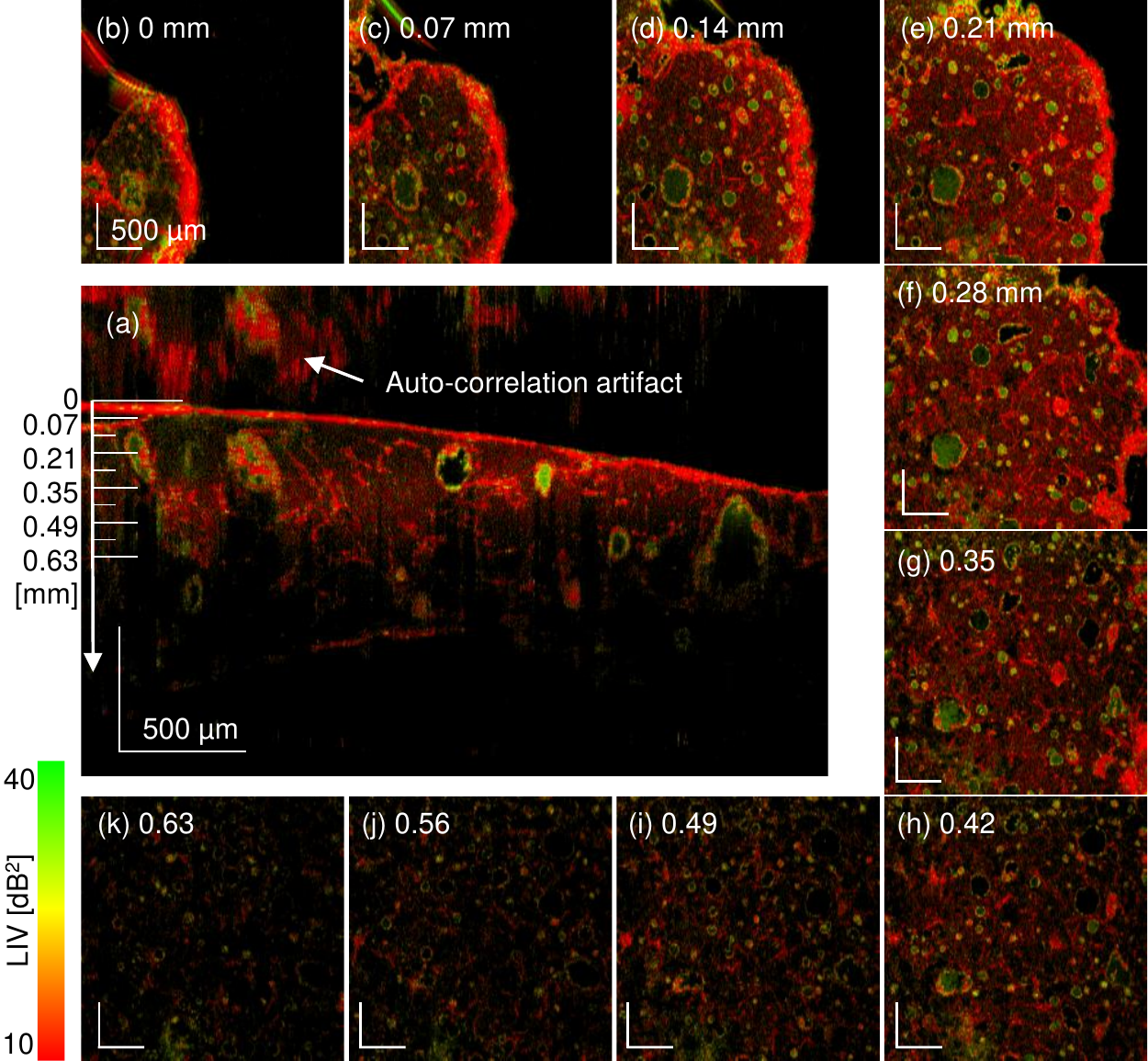}
	\caption{%
		The case of Day +7 bleomycin-model organoid.
	}
	\label{fig:SupDay7Bleomycin}
\end{figure}

\begin{figure}
	\centering\includegraphics{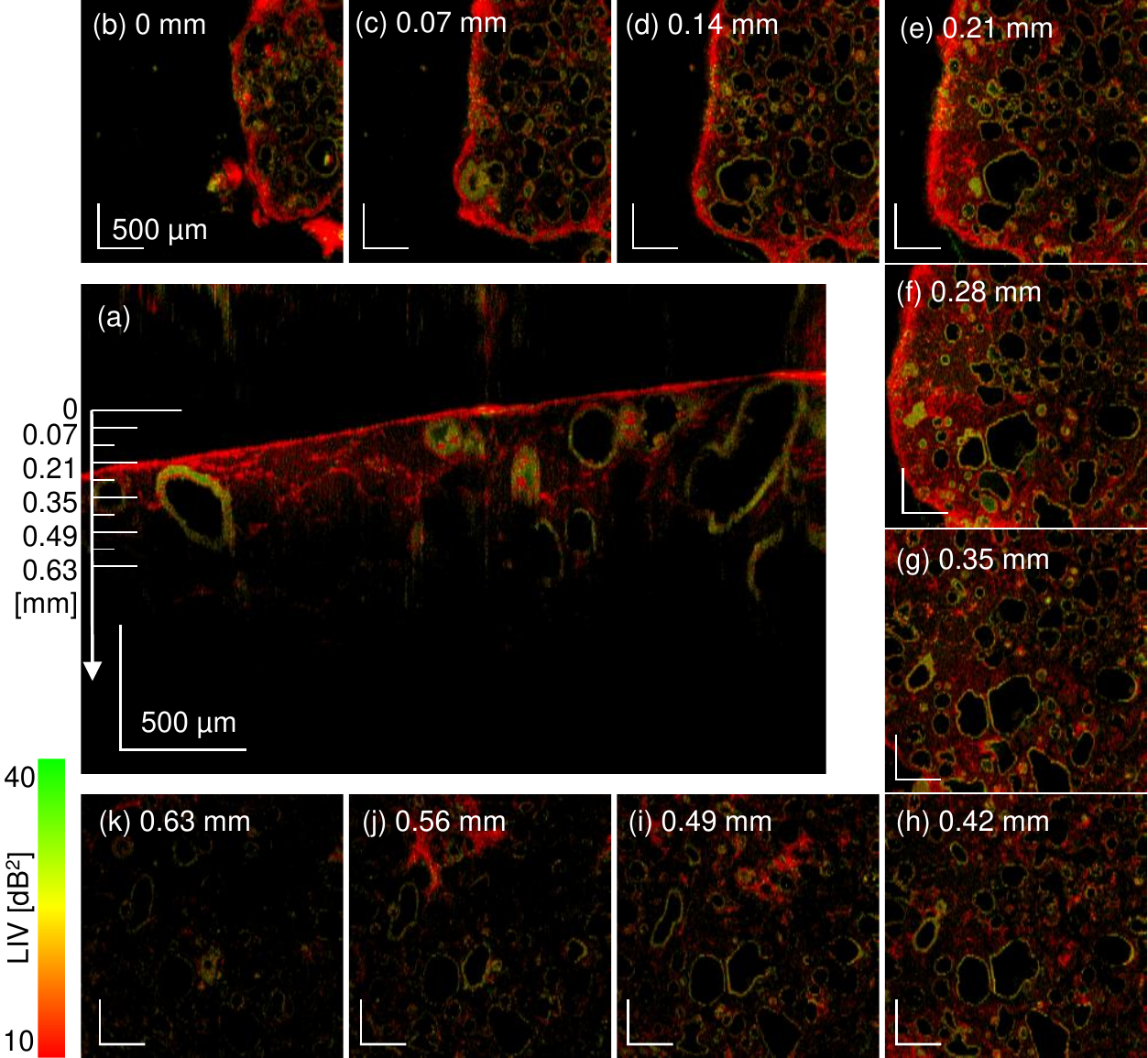}
	\caption{The case of Day +10 normal organoid.}
	\label{fig:SupDay10Normal}
\end{figure}

\begin{figure}
	\centering\includegraphics{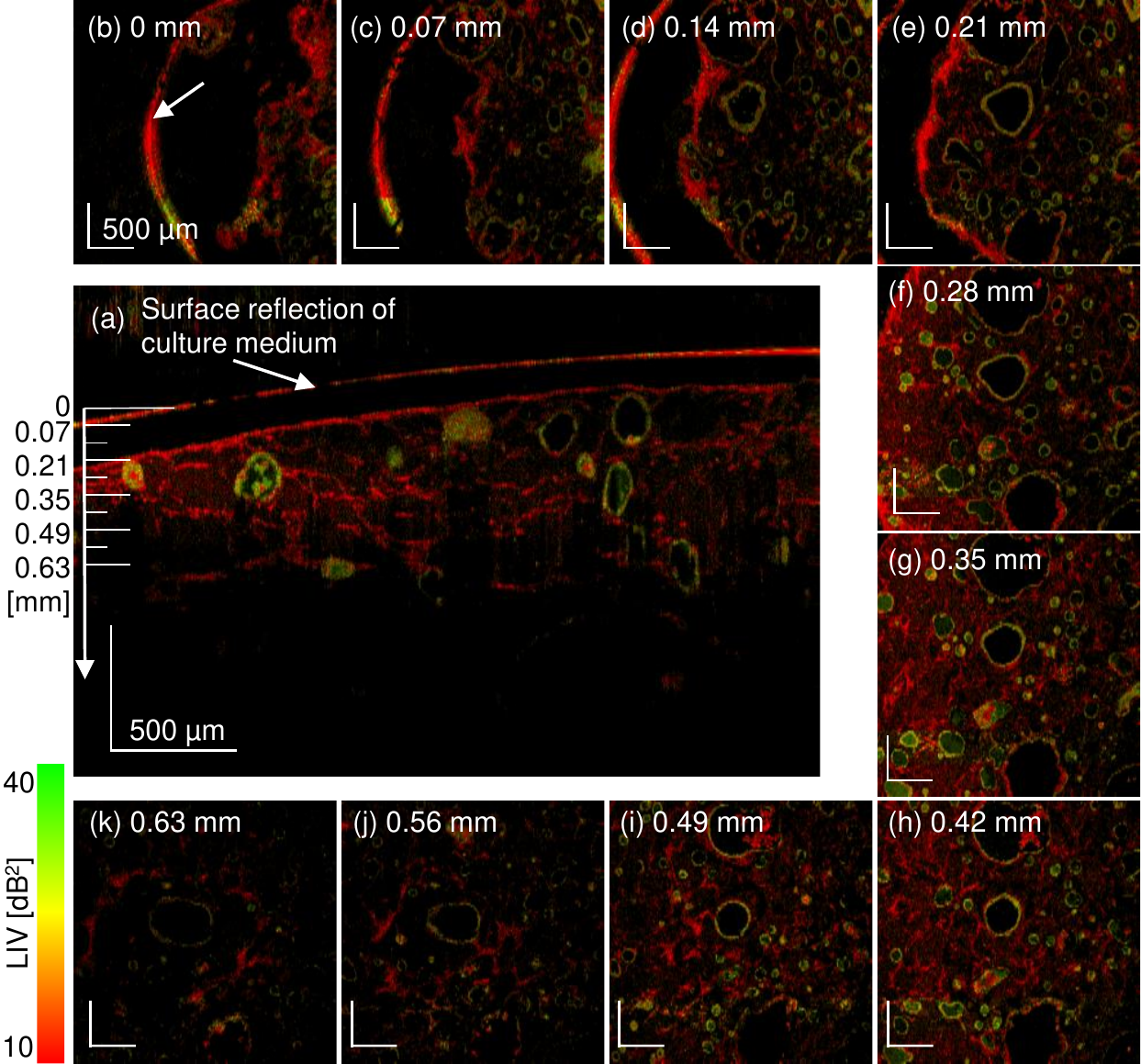}
	\caption{The case of Day +10 bleomycin-model organoid.
		The appearances indicated by white arrows are surface reflection of culture medium.}
	\label{fig:SupDay10Bleomycin}
\end{figure}

\begin{figure}
	\centering\includegraphics{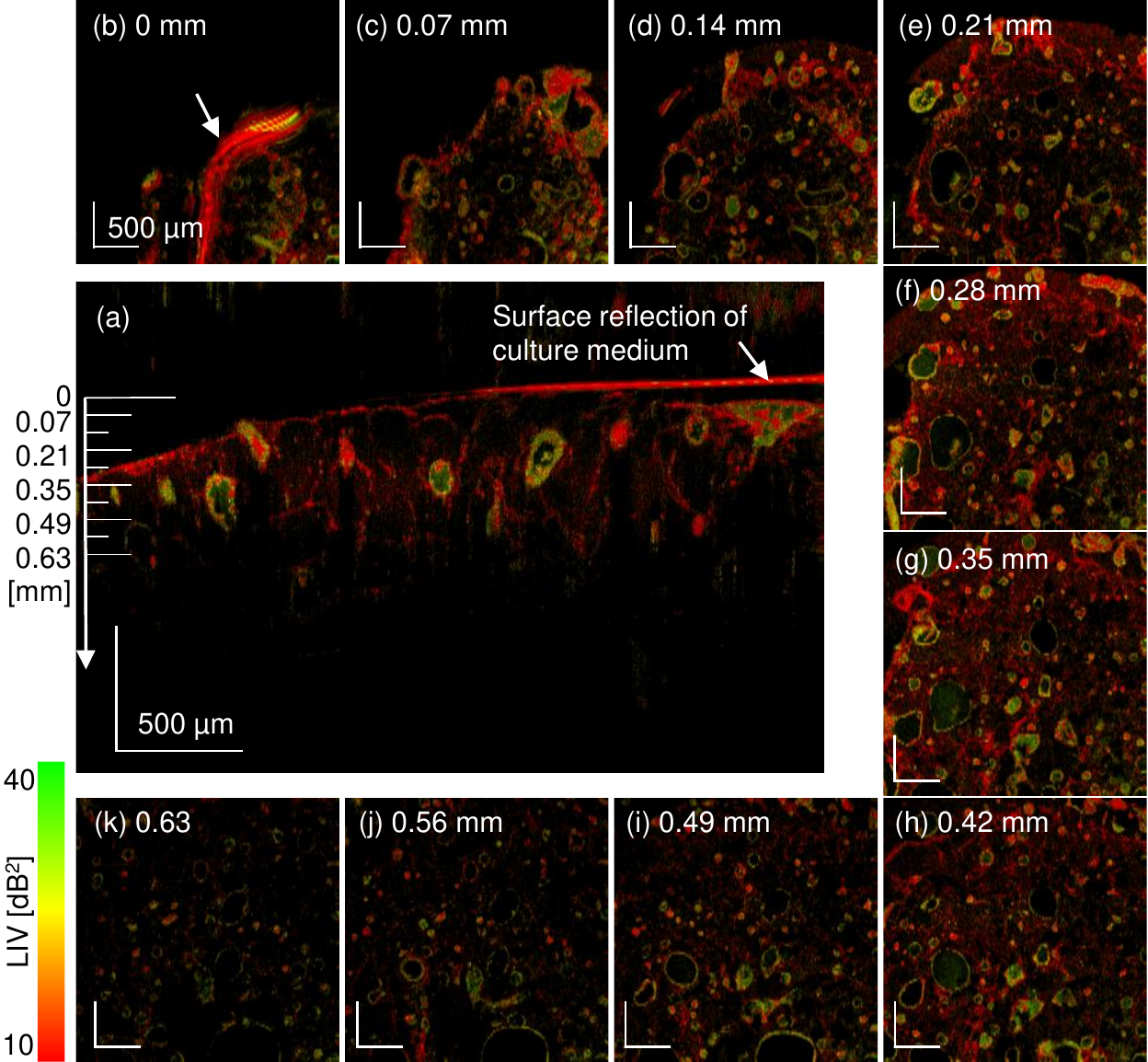}
	\caption{The case of Day +16 normal organoid.
		The appearances indicated by white arrows are surface reflection of culture medium.}
	\label{fig:SupDay16Normal}
\end{figure}

\begin{figure}
	\centering\includegraphics{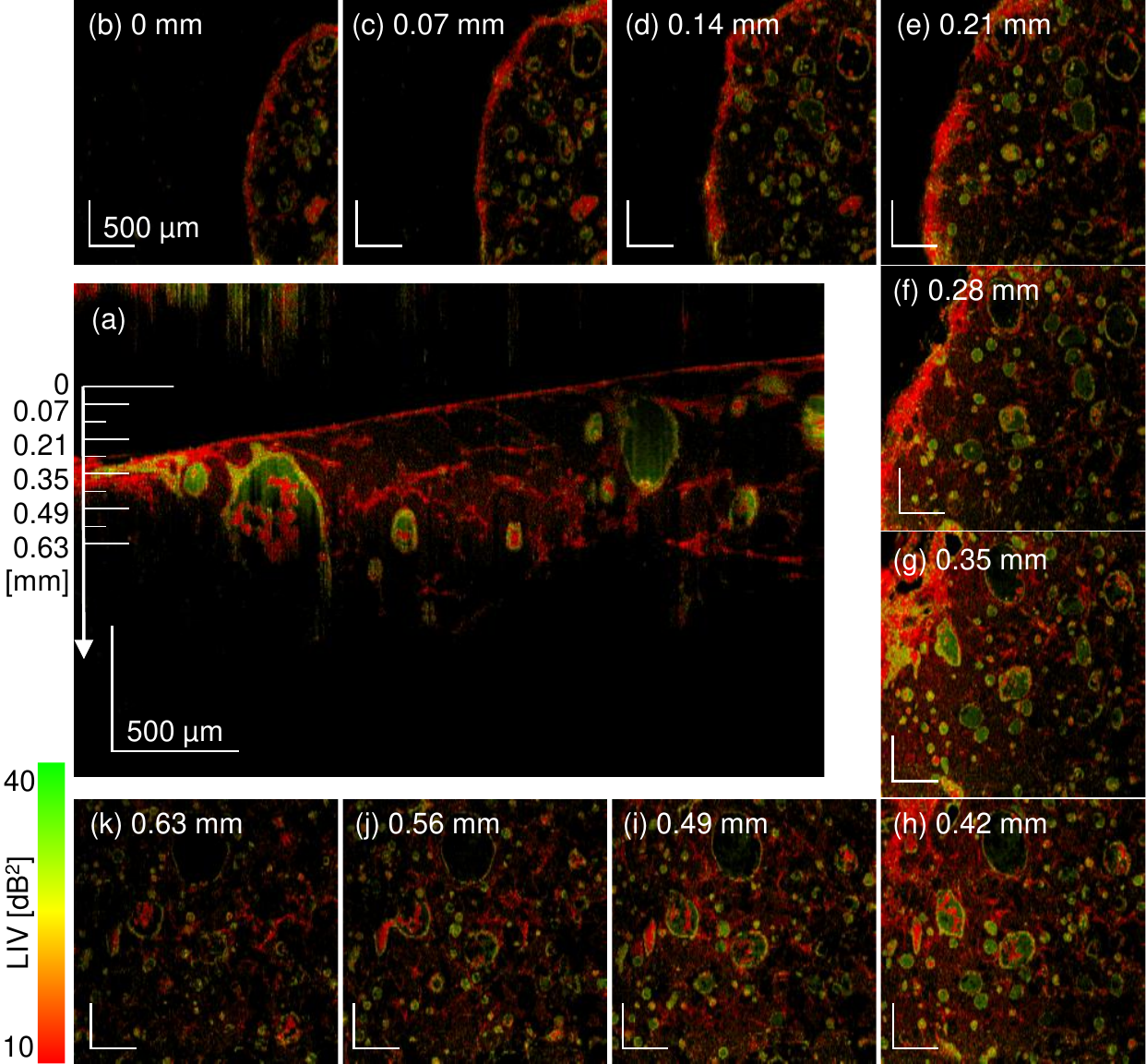}
	\caption{The case of Day +16 bleomycin-model organoid.}
	\label{fig:SupDay16Bleomycin}
\end{figure}

\begin{figure}
	\centering\includegraphics{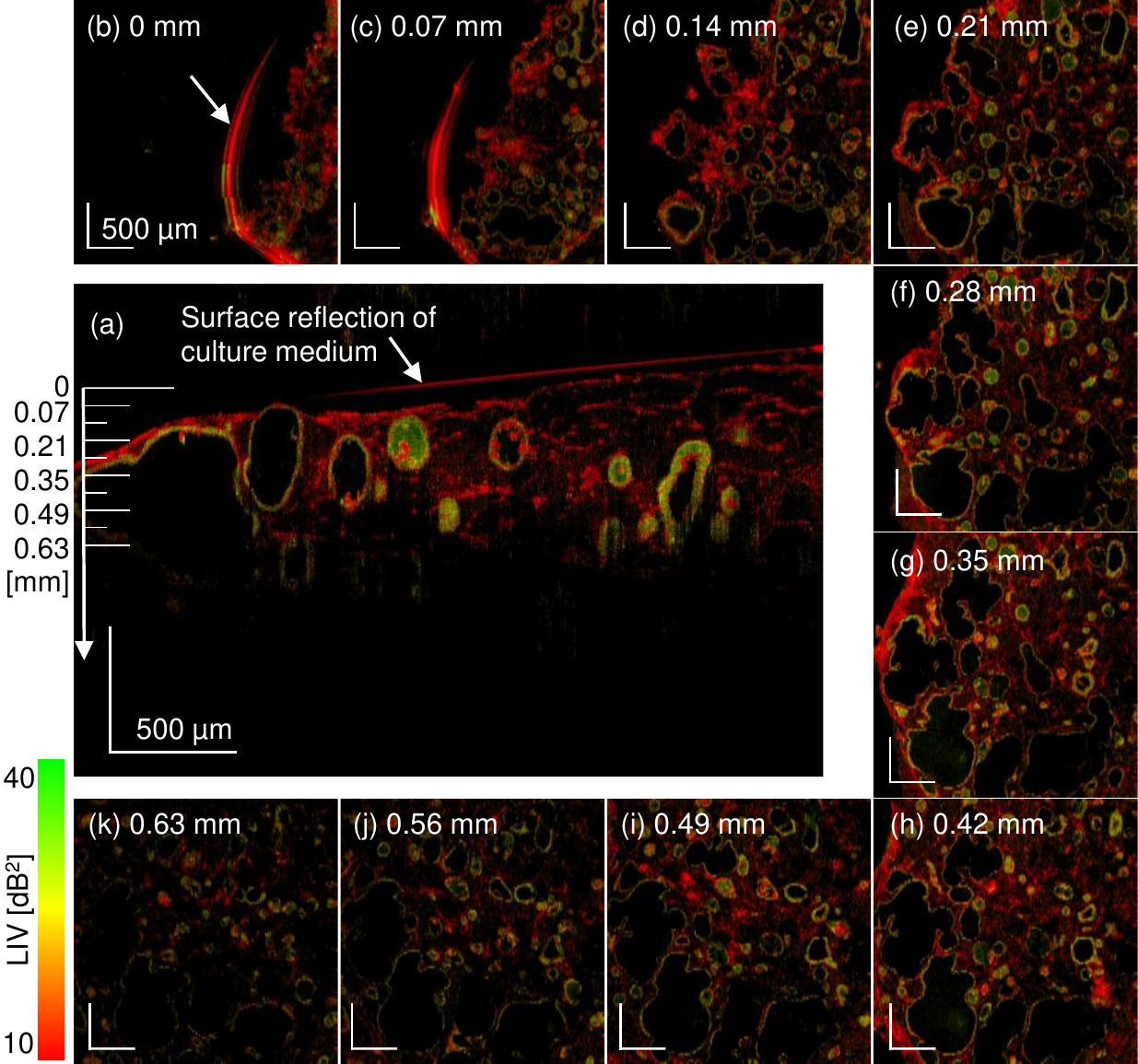}
	\caption{The case of Day +17 normal organoid.
		The appearances indicated by white arrows are surface reflection of culture medium.}
	\label{fig:SupDay17Normal}
\end{figure}

\begin{figure}
	\centering\includegraphics{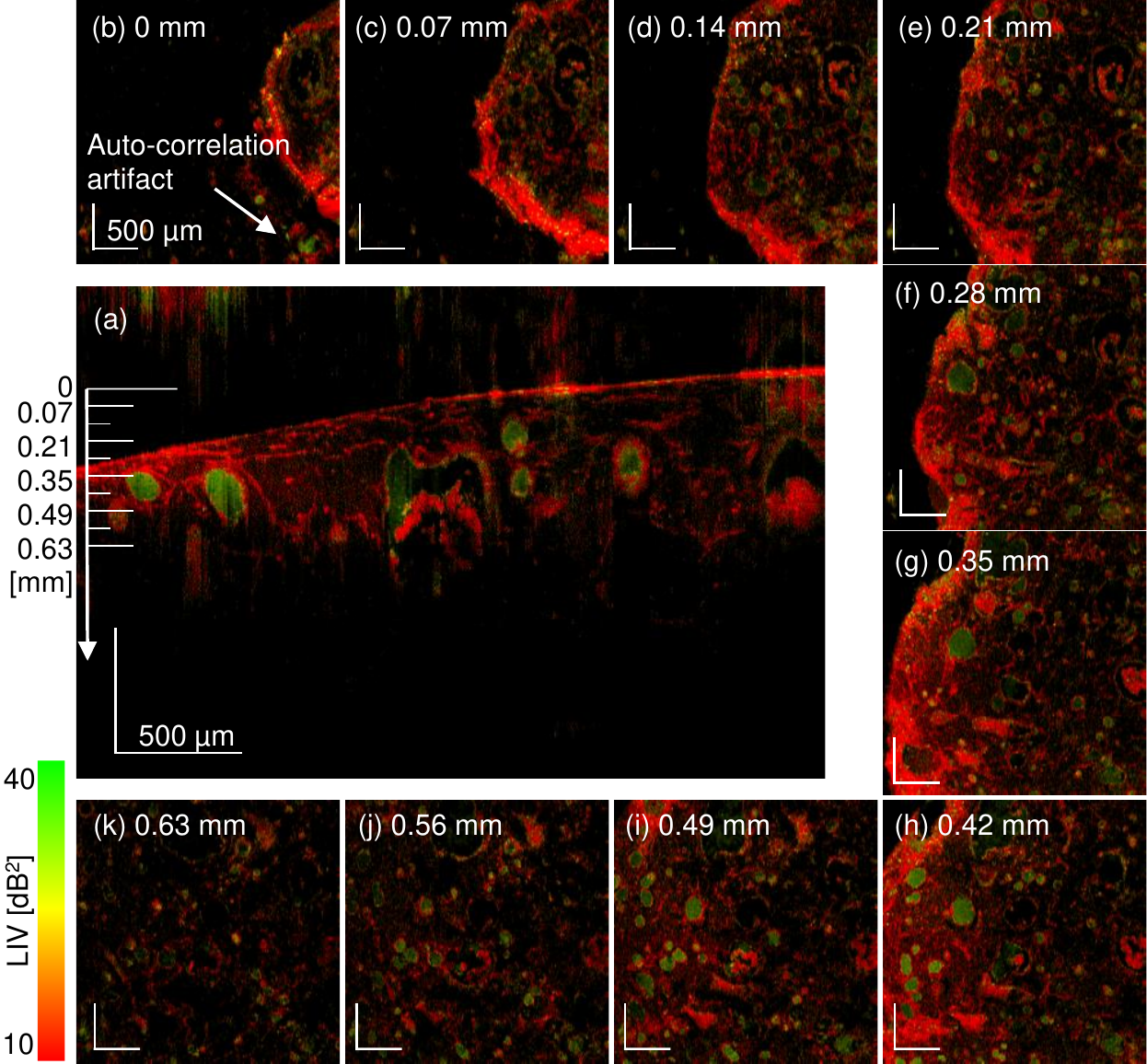}
	\caption{The case of Day +17 bleomycin-model organoid.
		The appearance indicated by a white arrow is an auto-correlation artifact.}
	\label{fig:SupDay17Bleomycin}
\end{figure}

\begin{figure}
	\centering\includegraphics{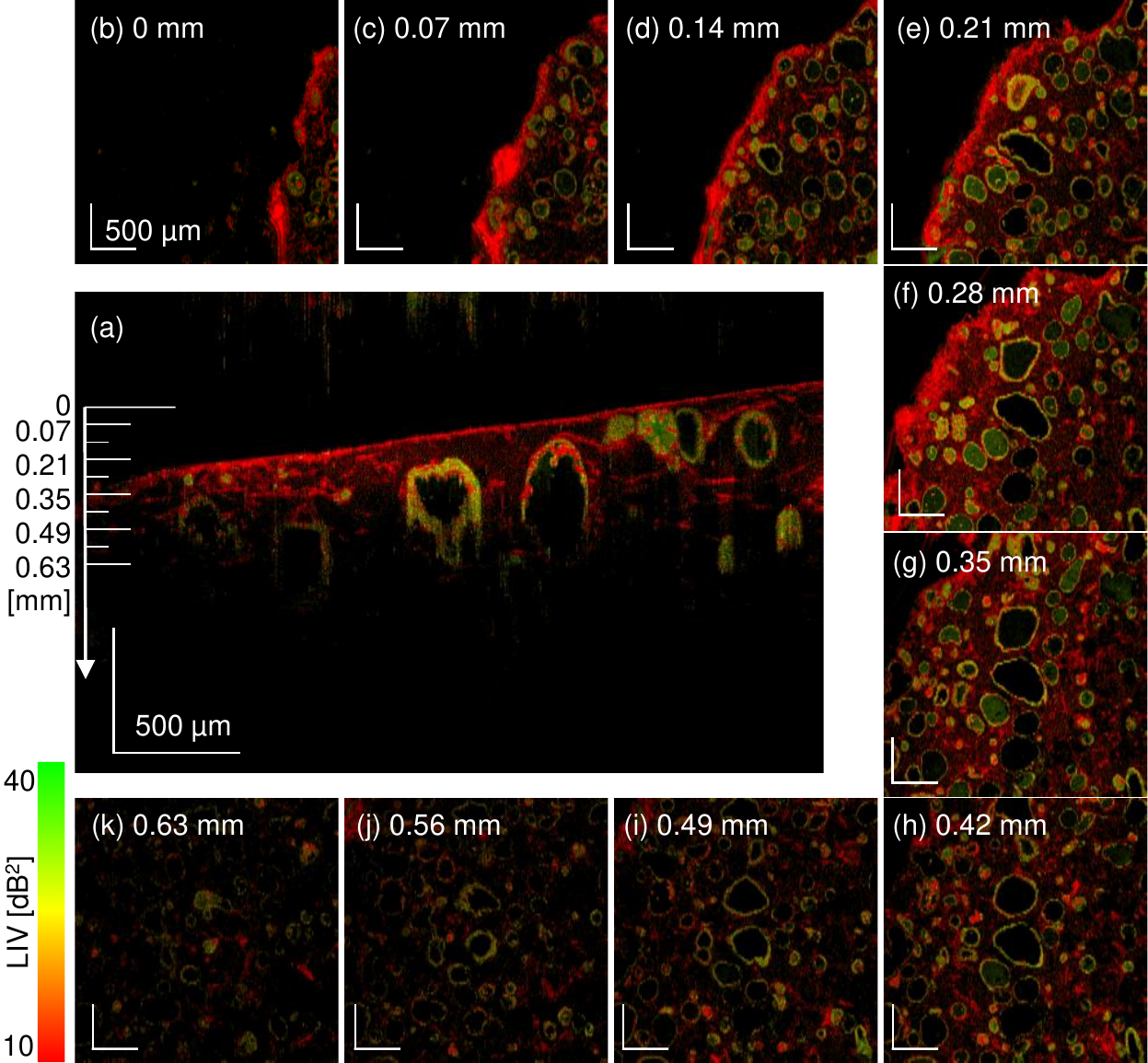}
	\caption{The case of Day +18 normal organoid.}
	\label{fig:SupDay18Normal}
\end{figure}

\begin{figure}
	\centering\includegraphics{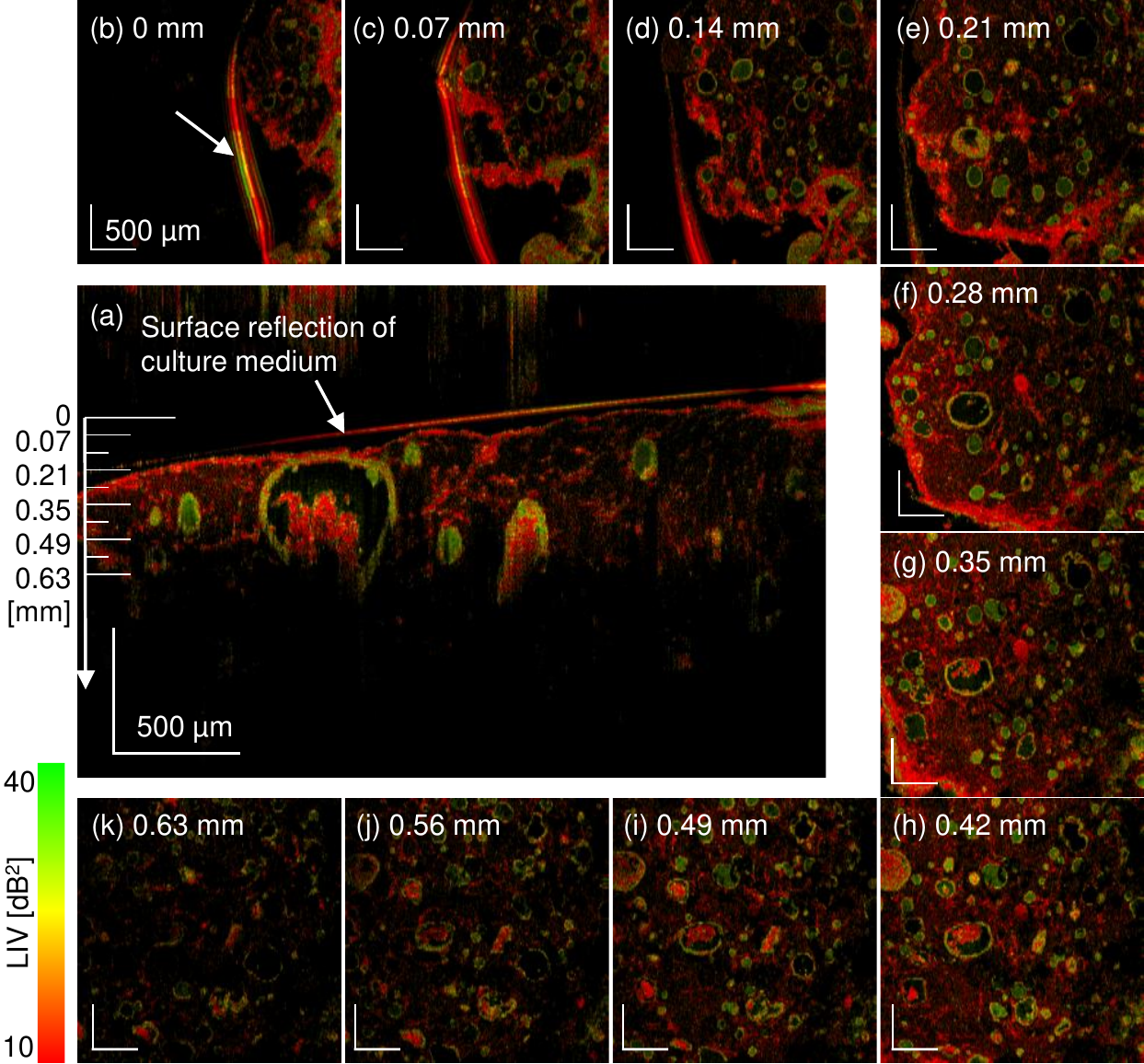}
	\caption{The case of Day +18 bleomycin-model organoid.
		The appearances indicated by white arrows are surface reflection of culture medium.}
	\label{fig:SupDay18Bleomycin}
\end{figure}

\end{document}